\documentstyle[aps,12pt,epsfig]{revtex}
\newcommand{\be}{\begin{equation}}
\newcommand{\ee}{\end{equation}}
\newcommand{\bea}{\begin{eqnarray}}
\newcommand{\eea}{\end{eqnarray}}
\newcommand{\beas}{\begin{eqnarray*}}
\newcommand{\eeas}{\end{eqnarray*}}
\newcommand{\bd}{\begin{displaymath}}
\newcommand{\ed}{\end{displaymath}}
\def\shiftleft#1{#1\llap{#1\hskip 0.04em}}
\def\shiftdown#1{#1\llap{\lower.04ex\hbox{#1}}}
\def\thick#1{\shiftdown{\shiftleft{#1}}}
\def\b#1{\thick{\hbox{$#1$}}}

\begin{document}
\baselineskip.9cm

\centerline{\Large Axial exchange currents and the spin content of the nucleon} 
%
%\centerline{\Large  in a nonrelativistic chiral quark model}  
%%%%%%%%%%%%%%%%%%%%%%%%%%%%%%%%%%%%%%%%%%%%%%
% Title too long, abstract explains what we do
%%%%%%%%%%%%%%%%%%%%%%%%%%%%%%%%%%%%%%%%%%%%%%
\vskip2cm

\centerline{\large D. Barquilla-Cano$^1$,  A. J. Buchmann$^2$ and E. Hern\'andez$^1$}
\vskip1cm 
\begin{center}
$^1$ Grupo de Fisica Nuclear, Facultad de Ciencias, Universidad de Salamanca  \\
Plaza de la Merced s/n, E-37008 Salamanca, Spain\\
$^2$ Institut f\"ur Theoretische Physik, Universit\"at T\"ubingen\\
Auf der Morgenstelle 14, D-72076 T\"ubingen, Germany \\
\end{center}
\vskip2cm
\centerline{\Large Abstract}

\noindent
In a chiral quark model where chiral symmetry is introduced via
a non-linear $\sigma$ model, 
we evaluate the  axial couplings $g_A(0)$, $g_A^8(0)$ and $g_A^0(0)$
related to the spin structure of the nucleon.
Our calculation
includes one-body and two-body axial current and pion absorption  
operators, which satisfy the Partial Conservation of Axial 
Current (PCAC) condition. While $g_A(0)$ is dominated 
by the one-body axial current we find significant corrections due to two-body 
axial exchange currents in $g_A^8(0)$ and $g_A^0(0)$. 
Interestingly, the axial current 
associated with gluon exchange reduces $g_A^8(0)$ from 1 to 0.6. Our analysis shows 
that the so-called ``proton spin crisis'' can be resolved in a  
constituent quark model in which PCAC is satisfied. 
Furthermore, we use the  PCAC constraint 
in order to determine the couplings of the $\eta$ and $\eta'$ mesons to nucleons.

\newpage

\noindent
\section{Introduction}
\label{section:introduction}
\nobreak
In a recent work~\cite{david} 
we have evaluated the axial form factors of
the nucleon, $g_A$, $g_P^{non-pole}$ and $g_T$ in the context of a 
chiral constituent quark model. Chiral symmetry was introduced via a 
linear $\sigma$-model with pions and $\sigma$-meson degrees of freedom. 
The model is consistent with PCAC, up to some order in the nonrelativistic 
expansion of the axial quark operators. The PCAC constraint for the non-pole
part of the axial current reads
%was formulated following Ref.~\cite{Ada91} as
\be
\label{pcac}
   [H,A^{non-pole,0}({\bf q})]-{\bf q} \cdot {\bf A}^{non-pole}({\bf q})=
   i\ \sqrt2\ f_{\pi}\ M^{\pi}({\bf q}),
\ee
where $H$ is the Hamiltonian of the three-quark system, 
including the center of mass motion, 
$\b{q}$ is the three-momentum 
transfer of the weak gauge boson,  $A^{non-pole,\,\mu}\ (\mu=0,1,2,3)$ stands 
for the non-pole part of the axial current operator,  $M^{\pi}$ is the 
pion absorption operator, and $f_{\pi}=92.4$ MeV is the empirical 
pion decay constant. Eq.(\ref{pcac}) plays a similar role in constraining
the form of the axial currents as the continuity equation for the electromagnetic
current.

One of the main consequences of the PCAC relation for the axial current 
is that axial coupling of the constituent quarks, $g_{Aq}$, is not equal to unity. 
Instead, it is related to the pion-quark coupling constant, $g_{\pi q}$, 
via a Goldberger-Treiman relation.
The second important consequence of the PCAC constraint is the necessity to 
include axial exchange currents into the quark model. The presence of two-body 
potentials in $H$ should be accompanied by two-body axial 
exchange currents and pion absorption operators if PCAC as formulated in  Eq.(\ref{pcac}) 
is to hold. The results of Ref.~\cite{david} show that the 
isovector axial coupling $g_A(0)$ of the nucleon 
is dominated by the one-body axial current.  
Gluon, pion, and scalar axial exchange currents,
although they are individually quite large, add
to an overall correction of less than $2\%$. 

In the present paper we are interested in the two remaining axial 
nucleon couplings, namely the flavor octet isosinglet coupling 
$g_A^8(0)$ and its flavor singlet counterpart $g_A^0(0)$ related to the
spin content of the nucleon. 
The spin fractions $\Delta q$ carried by the 
quarks of flavor $q$ are axial current matrix elements and  
can be evaluated in a constituent quark model. 
In the naive nonrelativistic quark model, which uses only one-body 
axial currents, one obtains $\Delta u=4/3$, $\Delta d=-1/3$ and $\Delta s=0$ and
from them $g_A(0)=5/3$ and $g_A^8(0)=g_A^0(0)=1$. 
On the other hand, the experimental numbers are: $\Delta u =0.84 \pm 0.02$,  
$\Delta d =-0.42 \pm 0.02$, and $\Delta s = -0.09 \pm 0.02$~\cite{abe} leading to 
$g_A(0)=1.26 \pm 0.04$,  $g_A^8(0)=0.60 \pm 0.08$ and 
$g_A^8(0)=0.33 \pm 0.06$. 

This marked disagreement between theory and experiment has often 
been interpreted as a severe shortcoming of the constituent 
quark model (``spin crisis''). 
Here, we show that once the PCAC violating 
%%%%%%%%%%%%%%%%%%%%%%%%%
%::::::::::::::::::::::::::::::::::::::::::::::::::::::::::::::::
% INSERTED ``PCAC violating'' 
%::::::::::::::::::::::::::::::::::::::::::::::::::::::::::::::::
and commonly used approximations: 
(i) axial charge of the constituent quarks $g_{Aq}$ is equal to unity
i.e., the same as for QCD quarks, 
(ii) neglect of two-body axial exchange currents,
%%%%%%%%%%%%%%%%%%%%%%%%%%%%
%which violate PCAC 
%::::::::::::::::::::::::::::::::::::::::::::::::::::::::::::::::::::::::::::::::::::::::::::::
% ``which violate PCAC'' at this point sounds as if the axial two-body currents violate PCAC.
% That was my reason for the above change in the March 1 version.
%::::::::::::::::::::::::::::::::::::::::::::::::::::::::::::::::::::::::::::::::::::::::::::::
%%%%%%%%%%%%%%%%%%%%%%%%%%%%%
are removed~\cite{david}, the calculated 
spin fractions are in good agreement with the experimental numbers.

The paper  is organized as follows.
In sect.~\ref{section:dis} we review the relation between the axial current matrix
elements and the spin fraction carried by the quarks as
determined in deep inelastic lepton-nucleon scattering and octet baryon decays.
Sect.~\ref{section:sigma}, reviews the non-linear $\sigma$-model of Manohar and Georgi,
which is the basis of the chiral quark model used in sect.~\ref{section:hamiltonian}.
The axial current operators are listed in sect.~\ref{section:currents}, and 
our results for the nucleon spin structure are
presented in sect.~\ref{section:results}. There we also calculate 
the $\eta NN$ and $\eta'NN$ coupling constants.
A summary of our results is presented in sect.~\ref{section:summary}.

\section{Spin structure of the nucleon}
\label{section:dis}
In naive nonrelativistic constituent quark models, in which the three constituent
quarks are in relative S-wave states, the nucleon spin is accounted for
by the sum of the constituent quark spins. 
Since the late 1980's we know from deep inelastic 
scattering experiments  that only a small fraction of the proton spin is 
carried by the spin of the QCD (or current) quarks,
and that there is a nonzero contribution to the nucleon spin 
arising from strange quark-antiquark pairs 
in the Dirac sea \cite{emc}. 
Recent experiments conclude that approximately $1/3$ of the spin of the
nucleon is carried by QCD quarks \cite{abe}. 
Calling $\Delta q$ the fraction
of the proton spin carried by quarks of flavor $q$, we have from Ref.~\cite{abe}
%%%%%%%%%%%%%%%%%%%%%%%%%%%%%%%%%%%%%%%%%%%%%%%%%%%%%%%%%%%%%%%%%%%%%%%%%%%%%%%%
\begin{eqnarray}
\label{deltaq}
\Delta u = \hspace{.35cm} 0.84 \pm 0.02,\hspace{.5cm} 
\Delta d =-0.42 \pm 0.02, \hspace{.5cm}
\Delta s = -0.09 \pm 0.02.
\end{eqnarray} 
%%%%%%%%%%%%%%%%%%%%%%%%%%%%%%%%%%%%%%%%%%%%%%%%%%%%%%%%%%%%%%%%%%%%%%%%%%%%%%%%
%%%%%%%%%%%%%%%%%%%%%%%%%%%%%%%%%%%%%%%%%%%%%%%%%%%%%%%%%%%%%%%%%%%%%%%%%%%%%%%
%$\Delta u =0.84 \pm 0.02$,  
%$\Delta d =-0.42 \pm 0.02$ and
%$\Delta s = -0.09 \pm 0.02$. 
%%%%%%%%%%%%%%%%%%%%%%%%%%%%%%%%%%%%%%%%%%%%%%%%%%%%%%%%%%%%%%%%%%%%%%%%%%%%%%%%%%%

%::::::::::::::::::::::::::::::::::::::::::::::::::::::::::::::::::::::::::::::::::::::
%Combined the two paragraphs into one.
%:::::::::::::::::::::::::::::::::::::::::::::::::::::::::::::::::::::::::::::::::::::: 
%

In a parton model description of the nucleon, 
the spin fractions of the total nucleon spin 
carried by the individual quark flavors are given as 
\begin{eqnarray}
\Delta q=\int_0^1 dx\, \big(\, q_+(x)+\bar{q}_+(x)-q_-(x)-
\bar{q}_-(x)\,\big).
\end{eqnarray}
Here, $q_{\pm}(x)$ is the probability of finding in the nucleon 
a QCD quark of flavor $q$ with momentum fraction $x$ of the 
total nucleon momentum having its helicity parallel (+) 
or antiparallel (-) to the proton helicity. 
The quark momentum distribution functions $q_{\pm}(x)$ 
can be further decomposed as 
\be
q_{\pm}(x)=q^{val}_{\pm}(x)+q^{sea}_{\pm}(x),
\ee 
where 
$q^{val}_{\pm}(x)$ ($q^{sea}_{\pm}(x)$)
is the contribution of the QCD valence (sea) quarks; 
$\bar{q}_{\pm}(x)$ stands for the antiquark distribution 
function in the sea.

In deep inelastic lepton-proton scattering one is measuring  
the spin-dependent structure function $g_1^p(x,Q^2)$ of the proton 
at high momentum transfers $Q^2$. The integral of this function
$\Gamma_1^p=\int_0^1 g_1^p(x,Q^2)\ dx$ is given in the scaling limit 
by the parton model result ~\cite{carlitz}
\be
\Gamma_1^p=\frac{1}{2}\sum_q e_q^2\Delta q = \frac{4}{18} \Delta u + \frac{1}{18} \Delta d
+\frac{1}{18} \Delta s  
\ee
%
%::::::::::::::::::::::::::::::::::::::::::::::::::::::::::::::::::::::::::::::::
%EXTENDED the equation in order to make the contribution of each flavor explicit. 
%:::::::::::::::::::::::::::::::::::::::::::::::::::::::::::::::::::::::::::::::::
%
%%%%%%%%%%%%%%%%%%
%\cite{bjorken} 
%%%%%%%%%%%%%%%%%%
where $e_q$ is the quark charge of flavor q.
The spin fractions can be expressed as axial vector current matrix elements 
\be
\langle\,p,s|\bar{q}\gamma_{\mu}
\gamma_5 q|p,s\rangle=\Delta q \, s_{\mu},
\ee 
where $s_{\mu}$ is the spin vector. 

Additional information 
on the spin fractions can be extracted from the weak semileptonic decay of 
octet baryons. From neutron $\beta$-decay one can extract the combination
$g_A^3(0)=\Delta u-\Delta d$ which is nothing but the nucleon axial coupling
constant $g_A(0)$. Also, from SU(3) flavor symmetry and experimentally known
hyperon $\beta$-decays 
one obtains the spin fraction combination $g_A^8(0)=\Delta u+\Delta d-2\Delta s$.
The two constants $g_A(0)$ and  $g_A^8(0)$ govern all octet baryon $\beta$-decays
in the flavor symmetry limit. The combined DIS and hyperon $\beta$-decay data
give values for each $\Delta q$ as quoted above.
While the use of SU(3) symmetry is not without controversy and has been 
sometimes severely critizied \cite{lipkin}, some authors think that SU(3) 
symmetry breaking effects are small, of the order of 10\%~\cite{flores}.

If the Dirac sea were SU(3) flavor symmetric, $g_A^8(0)$ would have the
interpretation as the fraction of the proton spin carried by valence QCD quarks,
whereas the total fraction of the proton spin carried by all QCD quarks is
given by the quantity $g_A^0(0)=\Delta u+\Delta d+\Delta s$. Using the 
numbers~\cite{abe} given above one obtains $g_A^0(0)=0.33\pm 0.06$, definitely
smaller than 1, which is the expectation according to the naive nonrelativistic 
quark model. This  is the origin of the so called ``spin crisis''. 
The experimental result may change somewhat. From the use of the flavor tagging 
technique at the  HERMES experiment it would be
possible to obtain the spin fractions directly from deep inelastic scattering 
experiments without relying on the SU(3) flavor symmetry assumption. Preliminary
results from the experiment show that $\Delta s>0$ in disagreement
with the previous determination~\cite{jackson}. This latter result has been
critizied on the  account that it would take  $g_A^8(0)$ out of any reasonable
bounds~\cite{leader}.

Before we calculate the axial nucleon couplings 
we will review the SU(3) extended version of the non-linear 
$\sigma$-model, which is appropriate for the evaluation of 
$g_A^8(0)$ and $g_A^0(0)$. 

%%%%%%%%%%%%%%%%%%%%%%%%%%%%%%%%%%%%%%%%%%%%%%%%%%%%%%%%%%%%%%%%%%%%%%%%%%%%%%%%%%%%%
% THE USE OF SU(3) FLAVOR SYMMETRY HERE HAS BEEN CRITIZED, eg. BY LIPKIN, PLB214(88)429.
% IT IS GOOD TO HAVE A COMMENT ON THE ERROR INTRODUCED BY THIS ASSUMPTION.
%%%%%%%%%%%%%%%%%%%%%%%%%%%%%%%%%%%%%%%%%%%%%%%%%%%%%%%%%%%%%%%%%%%%%%%%%%%%%%%%%%%%%%

%%%%%%%%%%%%%%%%%%%%%%%%%%%%%%%%%%%%%%%%%%%%%%%%%%%%%%%%%%%%%%%%%%%%%%%%%%%%%%%
%To this end, we will work in the SU(3) extended version of the non-linear $\sigma$-model, 
%which is appropriate for the evaluation of 
%$g_A^8(0)$ and $g_A^0(0)$. In the non-linear $\sigma$-model we 
%only have quarks, gluons and pseudo-scalar boson degrees of freedom.
%%%%%%%%%%%%%%%%%%%%%%%%%%%%%%%%%%%%%%%%%%%%%%%%%%%%%%%%%%%%%%%%%%%%%%%%%%%%%%%%%%%%%%%%%

\section{The non-linear $\sigma$ model}

%%%%%%%%%%%%%%%%%%%%%%%%%%%%%%%%%%%%%%%%%%%%%%
%
%THE EQUATIONS ARE STILL FROM AUGUST 2002.
% SOME CORRECTIONS IN THE EQUATIONS (MISSING INDICES) 
% HAVE BEEN MADE. MINOR CHANGES IN WORDING.
%
%%%%%%%%%%%%%%%%%%%%%%%%%%%%%%%%%%%%%%%%%%%%%%

\label{section:sigma}

Let us start with the effective Lagrangian
for the non-linear $\sigma$ model, which lends support to the notion 
of constituent quarks interacting via gluon and pseudoscalar 
boson exchange~\cite{manohar}. The effective Lagrangian is given by
\bea
\label{lagrangian}
{\cal L}=\overline{\Psi}i\gamma^{\mu}(D_{\mu}+{\cal V}_{\mu})
\Psi -M \overline{\Psi}\Psi+ g_{Aq}\overline{\Psi}\gamma^{\mu}\gamma_5{\cal A}_{\mu}
\Psi+\frac{f^2_{\pi}}{4}\ tr \left(\partial^{\mu}\Sigma^{\dagger}
\partial_{\mu}\Sigma\right)-\frac{1}{2}\ tr \left( F_{\mu\nu}\,
F^{\mu\nu}\right).
\eea
Here, $\Psi = (u, \, d, \, s)$ 
is the quark field and
$M=m_q {\bf 1}$  is the quark mass matrix prior to any explicit 
SU(3) breaking. 
Furthermore, $D_{\mu}$ is the color SU(3)$_C$ covariant derivative 
\be
D_{\mu}=\partial_{\mu}+i\,g\,  G_{\mu}, 
\qquad G_{\mu}=\frac{1}{2}\,\lambda_j^c \cdot G^c_{j,\mu},
\ee
where  $G_{\mu}$ is the gluon field and  the $\lambda_j^c $ are the 
color SU(3)$_C$ Gell-Mann matrices.
$F_{\mu\nu}$ is the gluon field strength tensor 
\be
F_{\mu\nu}=
\partial_{\mu}\,G_{\nu}-\partial_{\nu}\,G_{\mu}
+i\,g\,[G_{\mu},G_{\nu}].
\ee
The fields ${\cal V}_{\mu}$ and ${\cal A}_{\mu}$ appearing in
Eq.(\ref{lagrangian}) 
are given in terms of the matrix field $\xi$ derived from an octet of  pseudoscalar 
boson fields $\b{\Phi}$
\be
\label{boscur}
{\cal V}_{\mu}=\frac{1}{2}(\xi \partial_{\mu}\xi^{\dagger}
+\xi^{\dagger}\partial_{\mu}\xi), \qquad 
{\cal A}_{\mu}=\frac{i}{2}(\xi \partial_{\mu}\xi^{\dagger}
-\xi^{\dagger}\partial_{\mu}\xi).
\ee

The pseudo-scalar bosons $\b{\Phi}$ are the Goldstone bosons (GB)
associated with the spontaneous breaking of the 
$SU(3)_V\times SU(3)_A$ chiral symmetry. 
Following Ref.~\cite{manohar} we describe their  
dynamics in terms of
a $3\times 3$ matrix field $\Sigma$ given by
\be
\label{sigma}
\Sigma=e^{i\b{\lambda}\cdot  \b{\Phi}/F}=\xi^2
\ee
where $\b{\lambda}$ is a vector whose components $\lambda_j$  are the eight 
Gell-Mann matrices of SU(3) flavor symmetry, and $\b{\Phi}$ stands for 
the eight GB fields $\Phi_j$, $j=1 \cdots 8$. Here, $i=1,2,3$ stands for the 
isotriplet $\pi$, $i=4, \cdots 7$ for the two doublets of $K$,
and $i=8$ for the $\eta$.
With this normalization $F$ is the GB decay constant. Working
in lowest order as we shall do 
\be
F=f_{\pi}=92.4\ {\rm MeV} .
\ee

%::::::::::::::::::::::::::::::::::::::::::::::::::
% COMBINED the two preceding paragraphs.
%:::::::::::::::::::::::::::::::::::::::::::::::::::

The matrix field 
\be
\label{xi}
\xi=e^{i\b{\lambda} \cdot  \b{\Phi}/2f_{\pi}}
\ee
transforms under $SU(3)_V\times SU(3)_A$ as
\be
\label{chitrabos}
\xi  \stackrel{\stackrel{SU(3)_V}{ }}{\longrightarrow}  V\xi V^{\dagger} \qquad 
\xi  \stackrel{\stackrel{SU(3)_A}{ }}{\longrightarrow}  A^{\dagger}\xi U^{\dagger}
= U\xi A^{\dagger}
\ee
where $V=\exp(-i \b{\lambda}\cdot \b{\alpha}_V/2)$ and 
$A=\exp(-i \b{\lambda}\cdot \b{\alpha}_A/2)$ are global 
transformations belonging to $SU(3)_V$ and  $SU(3)_A$ respectively. 
As for $U$, it is a unitary matrix field that depends
on the axial transformation $A$ and the $\b{\Phi}$ Goldstone boson fields.

On the other hand the quark field, $\Psi$, transforms as
\be
\label{chitraqua}
\Psi  \stackrel{\stackrel{SU(3)_V}{ }}{\longrightarrow}  V \Psi  \qquad
\Psi  \stackrel{\stackrel{SU(3)_A}{ }}{\longrightarrow}  U \Psi
\ee
while the gluon fields are invariant under both SU(3)$_V$ and SU(3)$_A$
transformations.

At this point a comment on the value of $g_{Aq}$ in the effective Lagrangian 
is in order.  Each term in Eq.(\ref{lagrangian}) is separately invariant under the 
chiral group SU(3)$_V\times$ SU(3)$_A$. This is why one can introduce 
an axial quark coupling $g_{Aq}\neq 1$ in the model without violating 
chiral symmetry. The fact that $g_{Aq}$ for constituent quarks can be different 
from unity is well founded, both from the phenomenological as well as the 
theoretical point of view. Weinberg~\cite{w1,w2} has shown
that while constituent quarks have no anomalous magnetic moments, 
their axial coupling may be considerably renormalized by the strong 
interactions. Explicit calculation in~\cite{w2} gives
\be
g_{Aq}^2=1-\frac{m_q^2}{8\pi^2f_{\pi}^2},
\ee 
which corresponds to a $8\%$ reduction of $g_{Aq}$.
Using Witten's large $N_C$ counting rules, one finds, that in contrast
to the magnetic moment of the quarks, corrections to $g_{Aq}$ appear
already in order $N_C^0$ \cite{andreas}.
Finally, investigation in the constituent quark structure in the Nambu and
Jona-Lasinio model of Ref.\cite{vogl}
shows that  $g_{Aq}\approx 0.78 $.

Explicit SU(3) breaking mass terms are introduced in the model as
\bea
\label{cqmt}
-\overline{\Psi}\ e^{-i \gamma_5 \b{\lambda} \cdot  \b{\Phi}/2f_{\pi}}
\ M_0\ e^{-i \gamma_5 \b{\lambda} \cdot  \b{\Phi}/2f_{\pi}}\ \Psi,
\eea
where $M_0=diag(m^0_{u},m^0_{d},m^0_{s})$ is the current quark mass matrix.
After an expansion in powers of
$1/f_{\pi}$ we get to dominant order
\bea
\label{ww}
-\overline{\Psi}\,M_0\,\Psi+\frac{i}{2f_{\pi}}\,\overline{\Psi}\gamma_5
\{\lambda_j, M_0\}\Psi\ \Phi_j
+\frac{1}{8f^2_{\pi}}\,\overline{\Psi}\,
\{\lambda_j,\{\lambda_k, M_0\}\}\Psi\,
\Phi_j\Phi_k .
\eea
The last term generates a mass term for the GB when the quark bilinear is
replaced by its vacuum expectation value, see for instance
Ref.~\cite{weinbergbook}.

When putting everything together with the Lagrangian in Eq.(\ref{lagrangian})
we get to dominant order
\bea
\label{lag2}
{\cal L}=&& \overline{\Psi}(i\gamma^{\mu}\partial_{\mu}-M-M_0)
\Psi + \frac{1}{2}\,\partial^{\mu}\Phi_j
\partial_{\mu}\Phi_j-\frac{1}{2}\,m^2_{\Phi_j}\Phi^2_j
-\frac{1}{2}\ tr \left( F_{\mu\nu}\,
F^{\mu\nu}\right)\nonumber\\
&&\hspace{-0.5cm} 
-\,g\overline{\Psi}\gamma^{\mu}\,G_{\mu}
\Psi 
+\frac{g_{Aq}}{2f_{\pi}}\,\overline{\Psi}\gamma^{\mu}\gamma_5\lambda_j
\Psi\ \partial_{\mu}\,\Phi_j
+\frac{i}{2f_{\pi}}\overline{\Psi}\gamma_5
\{\lambda_j, M_0\}\Psi\ \Phi_j\nonumber\\
&&\hspace{-0.5cm} -\frac{1}{4f^2_{\pi}}\,f_{jkl}\,\overline{\Psi}
\gamma^{\mu}\lambda_j\Psi\,\Phi_k \partial_{\mu}\Phi_l
+\frac{1}{8f^2_{\pi}}\overline{\Psi}\,
\{\lambda_j,\{\lambda_k, M_0\}\}\Psi\,
\Phi_j\Phi_k, 
\eea
where $M+M_0$ gives our constituent quark mass. The $f_{jkl}$ are the SU(3) 
structure constants
%%%%%%%%%%%%%%%%%%%%%%%%%%%%%%%%%%%%%%%%%%%%%%%%%%%%%%%%%%%%%%%%%%%%%%%%%%%%%%%%
%Now $M'=M+M_0=diag(m_q, m_q, m_s)$ is the constituent quark total mass matrix
%%%%%%%%%%%%%%%%%%%%%%%%%%%%%%%%%%%%%%%%%%%%%%%%%%%%%%%%%%%%%%%%%%%%%%%%%%%%%%%%%
and the $m_{\Phi_j}$ are the masses of the GB. Expressions for these masses in
terms of the current quark masses and the vacuum expectation value of the quark 
bilinear are easily obtained from Eq.(\ref{ww}) and
are given in Ref.\cite{weinbergbook}. We assume that the mass
terms  break SU(3) flavor symmetry while preserving SU(2)$_I$ isospin
symmetry. This means we shall take $m^0_u=m^0_d$. For this quantity one can  use 
the value 
%\bea
$m^0_u=m^0_d=\overline{m}\simeq 6$\ MeV,
%\eea
while $m^0_s$ can be determined now from the pion
and kaon mass to be 
%\bea
$m^0_s=\frac{m^2_{K}}{m^2_{\pi}}\, 2\overline{m}-\overline{m}\simeq 146$\, MeV.
%\eea 
The actual values are not so important in this calculation, where the total
constituent quark mass is the relevant quantity. 

In Eq.(\ref{lag2}) we have apart from the kinetic energy and mass
terms the gluon-quark interaction and  GB-quark interaction terms. 
These interaction Lagrangians provide the vertices needed to build 
the one-gluon exchange potential of Fig. 1a,
the one-boson exchange potentials of Fig. 1b, the one-body contribution to the 
GB absorption operator of Fig. 3a, the two-body contribution to the 
GB absorption operator due to gluon exchange of Fig. 3b, and the 
two-body GB absorption operator due to one-boson exchange of Fig. 3d.
In addition, the diagram in Fig. 3c corresponding to two-body GB
absorption due to the confinement interaction is obtained after 
the confinement potential of Fig. 1c is introduced in sect.~\ref{section:hamiltonian}.

In contrast to the linear $\sigma$ model, the coupling between the $u$ and $d$ 
quarks and
the Goldstone bosons is now predominantly pseudo-vector in nature. Neglecting the
pseudo-scalar term whose coupling constant is given by $\overline{m}/f_{\pi}\ll 1$
we obtain for the  
 SU(2)$_I$ sector 
\be
{\cal L}_{\pi q}=\frac{g_{Aq}}{2 f_{\pi}}\ \overline{\Psi}\gamma^{\mu}
\gamma_5\b{\tau}\Psi \partial_{\mu}\b{\pi},
\ee
with $\b{\tau}$ being the isospin Pauli matrix and $\b{\pi}$ the pion
field. Writing the coupling constant in the usual way as $g_{\pi q}/2m_q$
we recover the Goldberger-Treiman relation at the quark level for the $u$ and
$d$ quarks
\be
\label{gaq}
g_{Aq}=f_{\pi}\frac{g_{\pi q}}{m_q},
\ee
where $m_q$ is the $u$ and $d$ constituent quark mass for which we use a value
of $m_q=313$\,MeV.
The way $g_{\pi q}$ is fixed in Ref.\cite{paco} is unchanged here because 
for on-shell quarks the pseudo-vector and the pseudo-scalar quark-pion couplings
are equivalent. This means we recover our  result~\cite{david} 
\beas
g_{Aq}=0.774.
\eeas

\subsection{Vector and axial currents}
The vector currents that derive from the Lagrangian in Eq.(\ref{lag2}) and the
transformation properties of the fields are given by
\bea
\label{vectorc}
 V^{\mu}_j=\overline{\Psi}\gamma^{\mu}\frac{\lambda_j}{2}\Psi
+\frac{g_{Aq}}{2f_{\pi}}\,f_{jkl}\ \overline{\Psi}\gamma^{\mu}\gamma_5\lambda_l\Psi\ \Phi_k
+f_{jkl}\ \Phi_k\partial^{\mu}\Phi_l
\eea
Using the equations of motion of the fields one finds that to dominant order the 
divergence of the vector currents are
\bea
\label{divervec}
\partial_{\mu} V_j^{\mu}=-\frac{i}{2} \overline{\Psi}[\lambda_j,M_0]\Psi .
\eea
This shows that only the vector currents corresponding to  $j=1,2,3$  (isospin)
and $j=8$ (hypercharge) are conserved to that order.

From these vector currents we can build up the electromagnetic current. 
With the quark charge given as 
\be
e\ \biggl(\frac{1}{2}\lambda_3+\frac{\sqrt3}{6}\lambda_8\biggr)
\ee
the electromagnetic current will be
\be
J^{\mu}=e\ \biggl(\ V^{\mu}_3+\frac{1}{\sqrt3}V^{\mu}_8\biggr),
\ee
which is exactly conserved to all orders.

Turning now to the axial currents one gets  for the non-pole
part of those currents
\bea
\label{axialc}
A^{non-pole,\, \mu}_j=g_{Aq}\,\overline{\Psi}\gamma^{\mu}\gamma_5
\frac{\lambda_j}{\sqrt2}\Psi
+\frac{1}{\sqrt2\ f_{\pi}}\,f_{jkl}\ 
\overline{\Psi}\gamma^{\mu}\lambda_l\Psi\ \Phi_k.
\eea
The boson pole part of the axial current
$A_j^{GB-pole, \mu} = \sqrt{2} f_{\pi} \partial^{\mu} \Phi_j$,
which is completely determined by the Goldstone boson fields, is omitted.

The first term  
in Eq.(\ref{axialc}) provides the extra vertex needed for the one-body  (Fig. 2a),
the two-body gluon (Fig. 2b), and two-body confinement (Fig. 2c) axial exchange 
current operators, and the second term in $A^{non-pole,\, \mu}$ 
gives the one for the two-body axial current operator due to one-boson exchange (Fig. 2d). 

%+\sqrt2 f_{\pi}\partial^{\mu}\Phi_j
The divergence of the non-pole part of the axial current 
is given to dominant order by
\bea
\label{diverax}
\partial_{\mu} A^{non-pole,\, \mu}_j=-\sqrt2\, f_{\pi}\,(
\partial_{\mu}\partial^{\mu}+m^2_{\Phi_j})\, \Phi_j
+\frac{i}{\sqrt2}\, \overline{\Psi}\{\lambda_j,M_0\}\Psi
\eea
where once again the equation of motion has been used.
%For non-strange baryons we will only need the components $j=1,2,3$ and 
%$j=8$. In that case and in the absence of loops, 
The second term in Eq.(\ref{diverax}) comes from the pseudoscalar quark-GB
coupling that we shall neglect for the $u$ and $d$ sector. In its absence we
have
\bea
\label{pcacgb}
\partial_{\mu} A^{non-pole,\, \mu}_j \equiv -\sqrt2\, f_{\pi}\, (
 \partial_{\mu}\partial^{\mu}+m^2_{\Phi_j})\, \Phi_j
\hspace{1cm};\hspace{1cm} j=1,2,3,8.
\eea
This means we shall have exact PCAC for each term
in the axial current, both 
at the relativistic level, and  to the order we work
in the expansion in $\Phi/f_{\pi}$.
% This result  holds independent 
%of the actual value of the axial coupling of the constituent quarks $g_{Aq}$. 
%%%%%%%%%%%%%%%%%%%%%%%%%%%%%%%%%%%%%%%%%%%%%%%%%%%%%%%%%%%%%%%%%%%%%%%%%%%%%%%%%%%%%%%
%:::::::::::::::::::::::::::::::::::::::::::::::::::::::::::::::::::::::::::::::::::::::
%I don't get the point. Isn't the GT relation also a consequence of the non-linear 
%sigma model. Can you explain? 
%::::::::::::::::::::::::::::::::::::::::::::::::::::::::::::::::::::::::::::::::::::::
%That was not the case in our previous work~\cite{david}, where $g_{Aq}$ 
%different from unity was a necessary consequence of the PCAC constraint. 
%Here, chiral symmetry allows an arbitrary value for $g_{Aq}$. 
%%%%%%%%%%%%%%%%%%%%%%%%%%%%%%%%%%%%%%%%%%%%%%%%%%%%%%%%%%%%%%%%%%%%%%%%%%%%%%%%%%%%%%
% ARBITRARY? IN THE END THE GT HAS TO HOLD AND $g_{Aq}$ is fixed as before, isn't it? 
%%%%%%%%%%%%%%%%%%%%%%%%%%%%%%%%%%%%%%%%%%%%%%%%%%%%%%%%%%%%%%%%%%%%%%%%%%%%%%%%%%%%%%%
%We fix $g_{Aq}$ by imposing the empirical coupling of pions to the 
%constituent $u$ and $d$ quarks (see Eq.(\ref{gaq})).

%We will not consider  the GB pole pieces
%that contribute to the pseudoscalar form factor alone.
%
\section{The nonrelativistic chiral quark model}
\label{section:hamiltonian}
In this section we will briefly discuss 
the Hamiltonian of the chiral quark model, as motivated by the
non-linear $\sigma$-model of the preceding section,
 and the  wave function of the nucleon.

\subsection{The Hamiltonian}
We are interested only in non-strange baryons so that only $u$ and $d$ quarks are
needed.
The Hamiltonian for the internal motion of the three quarks in a baryon 
is given by:
\be
\label{hamiltonian}
H_{}=\sum_{j}\left( m_q+\frac{\b{p}_j^2}{2m_q}\right)-
\frac{\b{P}^2}{6m_q}+\sum_{j<k}\left(
\left( V^{conf} \right)_{j,k}+
\left( V^{g} \right)_{j,k}+
\left( V^{\pi} \right)_{j,k}+
\left( V^{\eta_8} \right)_{j,k}
\right) ,
\ee
$\b{p}_j$ is the momentum operator of the j-th quark, and
$\b{P}$ is the momentum of the center of mass of the three quark system
whose contribution to the kinetic energy is subtracted from the total Hamiltonian.
Apart from the confinement potential $\left(V^{conf}\right)$, the 
Hamiltonian includes
two-body potentials from one-gluon  $\left(V^{g}\right)$,
one-pion $\left(V^{\pi}\right)$, and one-$\eta_8$, (or $\Phi_8$),
 $\left(V^{\eta_8}\right)$
exchange. The latter potentials are obtained from the Feynman diagrams in Fig.1.
For the $\pi$ and $\eta_8$ meson exchange potentials we introduce a short 
distance regulator by means of the static vertex form factor
\be
\label{ff}
F({\bf k}^2)=\left( \frac{\Lambda^2}{\Lambda^2+{\bf k}^2}\right)^{1/2}.
\ee
Here, $\bf{k}$ is the three-momentum of the exchanged meson and $\Lambda$
is the cut-off parameter. In coordinate space this leads to a very 
simple form for the potential where a second Yukawa 
term with a fictitious meson mass $\Lambda$ appears.
The actual expressions are given by
\bea
&& \hspace{-1.5cm} \left( V^{\pi} \right)_{j,k}= 
\left(\frac{g_{Aq}}{2f_{\pi}}\right)^2
\frac{1}{4\pi}
\frac{\Lambda^2_{\pi}}{\Lambda^2_{\pi}-m^2_{\pi}}\
\b{\tau}_j \cdot \b{\tau}_k\ 
(\b{\sigma}_j \cdot {\bf \nabla}_r)\
(\b{\sigma}_k \cdot {\bf \nabla}_r)
\left( \frac{e^{-m_{\pi}r}}{r}
 - \frac{e^{-\Lambda_{\pi}r}}{r}
\right)  \\
&& \hspace{-1.5cm} \left( V^{\eta _8} \right)_{j,k}=
 \left(\frac{g_{Aq}}{2f_{\pi}}\right)^2
\frac{1}{4\pi}
\frac{\Lambda^2_{\eta_8}}
{\Lambda^2_{\eta_8}-m^2_{\eta_8}}\
\frac{1}{3}\ (\b{\sigma}_j \cdot {\bf \nabla}_r)\
(\b{\sigma}_k \cdot {\bf \nabla}_r)
\left( \frac{e^{-m_{\eta_8} r}}{r} - \frac{e^{-\Lambda_{\eta_8} r}}{r}
\right) 
\eea 
where ${\bf r}_j$, $\b{\sigma}_j$ and $\b{\tau}_j$
are the position, spin and isospin operators of the j-th quark. The relative 
coordinate ${\bf r}$ is given by 
${\bf r}={\bf r}_j-{\bf r}_k$ and
$r=|{\bf r}|$. The factor $1/3$ in the $\eta_8$ potential comes from the
normalization of the SU(3) flavor matrix $\lambda_8$.
We will take the cut-off parameters  equal 
\bea
 \Lambda_{\eta_8}=\Lambda_{\pi}=\Lambda .
\eea
The value of the cut-off $\Lambda$ is fixed in Ref.\cite{paco}, 
by fitting the size of  the $q\bar{q}$ component  of the pion, assumed to
be of 0.4 fm as given in Ref.\cite{weise}. A value of
$\Lambda=4.2$ fm$^{-1}$ results.
For the masses we use $m_{\pi}=139$ MeV, and assuming no octet-singlet
mixing the Gell-Mann-Okubo mass formula 
%%%%%%%%%%%%%%%%%%%%%%%%%%%%%%%%%%%%%%%%%
%\cite{go} 
%%%%%%%%%%%%%%%%%%%%%%%%%%%%%%%%%%%%%%%%%
gives $m_{\eta_8}^2=\frac{1}{3}(4m_K^2-m_{\pi}^2)$ from where $m_{\eta_8}\approx 565$ MeV 
is obtained, which is very close to the mass of the physical $\eta$ meson.

The chiral quark model also contains the one-gluon exchange potential introduced by 
de R\'ujula et al. \cite{rujula}, and originally used to
% explain certain 
%regularities in 
explain
the spectrum of excited baryon states \cite{ik,ik2}.
For this gluon exchange potential we use
\bea
%\hspace{-1cm}
  \left( V^{g} \right)_{j,k}= \frac{\alpha_s}{4} 
\b{\lambda}^c_j \cdot \b{\lambda}^c_k
\Biggl\{\frac{1}{r}-\frac{\pi}{m_q^2}\left( 1+\frac{2}{3}\ \b{\sigma}_j 
\cdot \b{\sigma}_k \right) \delta ({\bf r})-
\frac{1}{4 m_q^2} \biggl( 
3\ (\b{\sigma}_j\cdot\hat{{\bf r}})\ 
(\b{\sigma}_k\cdot\hat{{\bf r}}) - \b{\sigma}_j\cdot\b{\sigma}_k\biggr)
\frac{1}{r^3}
\Biggl\},
\eea
where $\alpha_s$ is the effective quark-gluon coupling
constant that we will take as a free parameter, $\b{\lambda}^c$ are the 
$SU(3)$ color matrices and  $\hat{{\bf r}}={\bf r}/{r}$. Following 
the argumentation of Isgur and Karl~\cite{ik} we do not consider 
the spin-orbit terms in the calculation because it is known that such terms 
would heavily distort the spectrum of the negative parity baryons.

\par
The constituent quarks are confined by a long-range, spin-independent, scalar
two-body potential. For convenience a harmonic oscillator potential is often used
\be
\left( V^{conf}\right)_{j,k}=-a\ \b{\lambda}^c_j\cdot \b{\lambda}^c_k\ 
r^2.
\ee
However, lattice calculations suggest that a linear radial function
is more realistic. A linear confinement which a larger distances is 
screened by quark-antiquark
pair creation is found in some lattice calculations.
%%%%%%%%%%%%%%%%%%%%%%%%%%%%%%%%%%%%%%%%%%%%%%%%%%%%%%%%%%%%%%%%%%%%%%%%%% 
%The effect of color-
%screening potentials on the baryon spectrum has been investigated by Zhang 
%et al. \cite{zhang}.
%In this work our wave functions will contain configuration mixing.
%%%%%%%%%%%%%%%%%%%%%%%%%%%%%%%%%%%%%%%%%%%%%%%%%%%%%%%%%%%%%%%%%%%%%%%%%%%
It is known that  a pure h.o. potential without
any anharmonicity cannot reproduce the baryon mass spectrum \cite{ik,ik2},
 therefore we will use here a color screened  confinement potential of the form
\be
\label{pcc}
\left( V^{conf}\right)_{j,k}=-a\ \b{\lambda}^c_j\cdot \b{\lambda}^c_k
\left(1-e^{-\mu r}\right) +C,
\ee
where $a$, $\mu$ and $C$ are free parameters.
\subsection{The nucleon wave function}
Even though we work with the color screened confinement potential we will use
the harmonic oscillator basis to expand our wave functions. We shall restrict
ourselves to states up to $N=2$
excitation quanta. The nucleon wave function is thus given by a superposition 
of five h.o. states:                       
\be
\label{wf}
|N\rangle=a_{S_S}\ |S_S\rangle+a_{S_S'}\ |S_S'\rangle
+a_{S_M}\ |S_M\rangle+a_{D_M}\ |D_M\rangle
+a_{P_A}\ |P_A\rangle
\ee                                        
A complete description of the wave functions used in Eq.(\ref{wf}) can be found
in Ref.~{\cite{ik2}}. The mixing coefficients will be determined by diagonalization of the Hamiltonian
in Eq.(\ref{hamiltonian}) in this restricted h.o. basis. 

The free parameters of the model  are $\alpha_s$, $a$, $\mu$, $C$ and $b$,
where $b$ is the harmonic oscillator constant that appears in the radial part 
of the wave functions.  The nucleon spectrum alone 
does not provide sufficient constraints in order to fix the parameters uniquely. 
For that reason we include also the $\Delta$ spectrum and the low energy  
electromagnetic nucleon properties, such as the magnetic moments and the charge and
magnetic radii in order to fix the parameters. For details see Ref.~\cite{meyer}.
%%%%%%%%%%%%%%%%%%%%%%%%%%%%%%%%%%%%%%%%%%%%%%%%%%%%%%%%%%%%%%%%%%%%%%
%The $\Delta$ wave functions are expanded as
%\be
%\label{wfd}
%|\Delta\rangle=b_{S_S}\ |S_S\rangle_{\Delta}+b_{S_S'}\ |S_S'\rangle_{\Delta}
%+b_{D_S}\ |D_S\rangle_{\Delta}+b_{D_M}\ |D_M\rangle_{\Delta}
%\ee                                        
%A complete 
%description of the wave functions used in
%Eqs.(\ref{wf},\ref{wfd})
% can be found in Ref.\cite{ik2}.
%%%%%%%%%%%%%%%%%%%%%%%%%%%%%%%%%%%%%%%%%%%%%%%%%%%%%%%%%%%%%%%%%%%%%%%%%%%%%%% 
The model parameters  are shown in Table~ \ref{table:parameters} 
while in Table~ \ref{table:admixtures} we give the
admixture coefficients for the nucleon wave function. The $S_S$ component 
clearly dominates.

\section{The axial current operators}
\label{section:currents}
%%%%%%%%%%%%%%%%%%%%%%%%%%%%%%%%%%%%%%%%%%%%%%%%%%%%%%%%%%%%%%%%%%%%%%%%
%We  obtain the axial operators from the Feynman diagrams of Fig. 2.
%We include the one-body or impulse contribution of Fig. 2a and 
%the two-body pieces corresponding to one-gluon exchange in Fig. 2b,
%confinement interaction in Fig. 2c, and Goldstone-boson exchange in
%Fig. 2d. For non-strange baryons only pions contribute in Fig. 2d.
%In Fig. 3(a-d) we show the Feynman diagrams for the corresponding 
%Goldstone-boson absorption processes.
%%%%%%%%%%%%%%%%%%%%%%%%%%%%%%%%%%%%%%%%%%%%%%%%%%%%%%%%%%%%%%%%%%%%%%%%%

Using the vertices extracted from Eqs.(\ref{lag2}, \ref{axialc}) it can be
readily shown that each separate contribution in Figs. 2-3 satisfies 
in momentum space 
\bea
\label{pcacms}
q_{\mu}\ A^{non-pole,\, \mu}_{j}= i\ \sqrt2\ f_{\pi}\ M^{GB}_j
\eea
for the components $j=1,2,3,8$ of the current and for 
non-strange quarks. Here, $q$ is the four-momentum transfer, 
$A^{non-pole,\,\mu}_j$ is the axial current amplitude
corresponding to Fig.2(a-d), 
and $M^{GB}_j$  the Goldstone boson absorption amplitude of Fig.3(a-d).
This result is a direct consequence of Eq.(\ref{pcacgb}).

Let us now give the explicit expressions for the different axial current operators
that correspond to the nonrelativistic reduction 
%%%%%%%%%%%%%%%%%%%%%%%%%%%%%
%of the currents derive from
%%%%%%%%%%%%%%%%%%%%%%%%%%%%
of the Feynman diagrams in Fig.2(a-d). Even though
we will only need the spatial part of these operators we give
here, for completeness, also the expressions for the zeroth component, i.e.,
for the axial charge operators. In the rest of this section we drop the
{\it non-pole} affix of the axial current operators in the understanding that we will always refer to its non-pole
part.
The one-body isovector axial current operators ($j=1,2,3$) are  
\bea
\label{one-body isovector}
A^0_{j, imp}&=& g_{Aq}\ \sum_k\ \frac{\b{\tau}_k^{(j)}}{\sqrt2}
\ e^{i{\bf q}\cdot{\bf r}_k}\frac{1}{2m_q}\ \b{\sigma}_k\cdot({\bf q}
+2{\bf p}_k)\nonumber\\
{\bf A}_{j, imp}&=& g_{Aq}\ \sum_k\ \frac{\b{\tau}_k^{(j)}}{\sqrt2}
\ e^{i{\bf q}\cdot{\bf r}_k}\ \b{\sigma}_k .
\eea

In order to generalize the isovector axial current to the flavor octet isosinglet
axial current we replace in the axial vertex $\b{\tau}_k^{(j)}/\sqrt{2}$ 
by $\sqrt{3}\,\lambda_k^8$ to get
for the $j=8$  case
\bea
\label{a8i}
A^0_{8, imp}&\equiv& g_{Aq}\ \sum_k\ 
\ e^{i{\bf q}\cdot{\bf r}_k}\frac{1}{2m_q}\ \b{\sigma}_k\cdot({\bf q}
+2{\bf p}_k)\nonumber\\
{\bf A}_{8, imp}&\equiv& g_{Aq}\ \sum_k\ 
\ e^{i{\bf q}\cdot{\bf r}_k}\ \b{\sigma}_k,
\eea
where we have taken  into account that nucleons are made of 
$u$ and $d$ valence quarks so that we only need the left upper corner of the
SU(3) flavor matrix $\lambda_8=1/\sqrt{3}\, diag(1, \, 1, \, -2)$ which is proportional
to the unit matrix.

%%%%%%%%%%%%%%%%%%%%%%%%%%%%%%%%%%%%%%%%%%%%%%%%%%%%%%%%%%%%%%%%%%%%%%%%%%%%
% THE SIGNS OF THESE CURRENTS DO NOT AGREE WITH THE SIGNS IN THE PCAC PAPER
%   The signs are correct.In the other paper we took the -(+1) component
%   apprpriate for the n p transition
%%%%%%%%%%%%%%%%%%%%%%%%%%%%%%%%%%%%%%%%%%%%%%%%%%%%%%%%%%%%%%%%%%%%%%%%%%%%

As in the derivation of the two-body potentials, we consistently neglect 
all non-local terms in the exchange currents. 
For the one-gluon exchange operators  we then obtain for $j=1,2,3$
\bea
A^0_{j, g}&=& g_{Aq}\sum_{k<l}
\frac{\alpha_s}{8m_q^2}\ \b{\lambda}_k^c\cdot \b{\lambda}_l^c\
\Biggl\{\ \frac{-\b{\tau}_k^{(j)}}{\sqrt2} e^{i{\bf q}\cdot {\bf r}_k}\ 
(\b{\sigma}_k\times \b{\sigma}_l)\cdot {\bf r}
+(k\leftrightarrow l)\Biggr\}\frac{1}{r^3}\nonumber\\[.3cm] 
{\bf A}_{j, g}&=& g_{Aq}\sum_{k<l}
\frac{\alpha_s}{16m_q^3}\ \b{\lambda}_k^c\cdot \b{\lambda}_l^c\
\Biggl\{ \frac{\b{\tau}_k^{(j)}}{\sqrt2}\ \ e^{i{\bf q}{\bf r}_k}\Biggl[ 
\ \Biggl(-i(\b{\sigma}_k\cdot {\bf r})\ {\bf q}\nonumber \\
&&\ \ \ \ \ \ \ \ \ \ \ +\biggl( 3((\b{\sigma}_k+\b{\sigma}_l)\cdot 
\hat{{\bf r}})\ \ \hat{{\bf r}}
-(\b{\sigma}_k+\b{\sigma}_l)  \biggr)\ \Biggr)\frac{1}{r^3}\nonumber \\
&&\ \ \ \ \ \ \ \ \ \ \ +\frac{8\pi}{3}(\b{\sigma}_k+\b{\sigma}_l)\ 
\delta({\bf r})\Biggr]
+(k\leftrightarrow l)\Biggr\} 
\eea
and for the 8-th component
\bea
\label{a8g}
A^0_{8, g}&\equiv&- g_{Aq}\sum_{k<l}
\frac{\alpha_s}{8m_q^2}\ \b{\lambda}_k^c\cdot \b{\lambda}_l^c\
\Biggl\{\  e^{i{\bf q}{\bf r}_k}\ 
(\b{\sigma}_k\times \b{\sigma}_l)\cdot {\bf r}
+(k\leftrightarrow l)\Biggr\}\frac{1}{r^3}\nonumber\\[.3cm] 
{\bf A}_{8, g}&\equiv& g_{Aq}\sum_{k<l}
\frac{\alpha_s}{16m_q^3}\ \b{\lambda}_k^c\cdot \b{\lambda}_l^c\
\Biggl\{ \ \ e^{i{\bf q}{\bf r}_k}\Biggl[ 
\ \Biggl(-i(\b{\sigma}_k\cdot {\bf r})\ {\bf q}\nonumber \\
&&\ \ \ \ \ \ \ \ \ \ \ +\biggl( 3((\b{\sigma}_k+\b{\sigma}_l)\cdot 
\hat{{\bf r}})\ \ \hat{{\bf r}}
-(\b{\sigma}_k+\b{\sigma}_l)  \biggr)\ \Biggr)\frac{1}{r^3}\nonumber \\
&&\ \ \ \ \ \ \ \ \ \ \ +\frac{8\pi}{3}(\b{\sigma}_k+\b{\sigma}_l)\ 
\delta({\bf r})\Biggr]
+(k\leftrightarrow l)\Biggr\} .
\eea
Now the confinement exchange contribution reads for $j=1,2,3$ 
\bea
A^0_{j,conf}& = & 0\nonumber \\[.3cm]
{\bf A}_{j, conf} &=& i g_{Aq}\frac{1}{4m_q^3}
\sum_{k<l}\ \Biggl\{
\frac{\b{\tau}_k^{(j)}}{\sqrt2}\ 
e^{i{\bf q}\cdot{\bf r}_k}\ \Biggl[
-i(\b{\sigma}_k\cdot{\bf q})\ {\bf q}
+({\bf q}\cdot\b{\nabla}_r)\ \b{\sigma}_k
-(\b{\sigma}_k\cdot\b{\nabla}_r)\ {\bf q}
\Biggr]\nonumber\\
&& \hspace{3cm} +(k\leftrightarrow l)\Biggr\}\ V^{conf}(r)\
\eea
and similarly for the 8-th component
\bea
A^0_{8,conf}& \equiv & 0\nonumber \\[.3cm]
{\bf A}_{8, conf} &\equiv& i g_{Aq}\frac{1}{4m_q^3}
\sum_{k<l}\ \Biggl\{
\ 
e^{i{\bf q}\cdot{\bf r}_k}\ \Biggl[
-i(\b{\sigma}_k\cdot{\bf q})\ {\bf q}
+({\bf q}\cdot\b{\nabla}_r)\ \b{\sigma}_k
-(\b{\sigma}_k\cdot\b{\nabla}_r)\ {\bf q}
\Biggr]\nonumber\\
&& \hspace{3cm} +(k\leftrightarrow l)\Biggr\}\ V^{conf}(r)\ .
\eea
Finally axial pion-pair exchange current operators  are given for $j=1,2,3$ by
\bea
\label{pion}
A^0_{j, \pi}&=& \frac{g_{Aq}}{2 f^2_{\pi}}\frac{1}{4\pi}
\frac{\Lambda^2}{\Lambda^2-m^2_{\pi}}\ \sum_{k<l}\ \Biggl\{
\frac{-(\b{\tau}_k\times \b{\tau}_l)^{(j)}}{\sqrt2}\ 
e^{i{\bf q}\cdot{\bf r}_k}\ \frac{1}{r}\ {\cal Y}_1(r)
\ \b{\sigma}_l\cdot{\bf r}+(k\leftrightarrow l)\Biggr\}\nonumber\\[.3cm]
{\bf A}_{j, \pi}&=& \frac{g_{A q}}{2 f^2_{\pi}}\frac{1}{2m_q}\frac{1}{4\pi}
\frac{\Lambda^2}{\Lambda^2-m^2_{\pi}}\  \ \sum_{k<l}\ \Biggl\{
e^{i{\bf q}\cdot{\bf r}_k}\ 
\frac{-(\b{\tau}_k\times \b{\tau}_l)^{(j)}}{\sqrt2}\
\Biggl[
\ (\b{\sigma}_k\times \b{\sigma}_l)
\ \frac{1}{r}\ {\cal Y}_1(r)\nonumber\\
&&\ \ \ \ \ \ \ \ + i(\b{\sigma}_k\times {\bf q})\ (\b{\sigma}_l\cdot{\bf r})
 \frac{1}{r}\ {\cal Y}_1(r) \nonumber\\
&&\ \ \ \ \ \ \ \ -\ 
(\b{\sigma}_l\cdot{\bf r})\ (\b{\sigma}_k\times{\bf r})
\ \frac{1}{r^2}\ {\cal Y}_2(r)
\Biggr] +(k\leftrightarrow l) \Biggr\}. 
%%%%%%%%%%%%%%%%%%%%%%%%%%%%%%%%%%%%%%%%%%%%%%%%%%%%%%%%%%%%%%%%%%%%%%%%%%%%%%%%%%%%%%%
%:::::::::::::::::::::::::::::::::::::::::::::::::::::::::::::::::::::::::::::::::::::::::
% This equation should also have number, whereas the next two equations do not need
% equation numbers.
%:::::::::::::::::::::::::::::::::::::::::::::::::::::::::::::::::::::::::::::::::::::::::
\eea
${\cal Y}_1(r)$  and 
${\cal Y}_2(r)$ are given by
\bea
&&{\cal Y}_1(r)= m^2_{\pi}\ Y_1(m_{\pi}r)-\Lambda^2\ Y_1(\Lambda r)
\nonumber\\
&&{\cal Y}_2(r)= m^3_{\pi}\ Y_2(m_{\pi}r)-\Lambda^3\ Y_2(\Lambda r) \nonumber
\eea
where
\bea
&&Y_1(x)=\frac{e^{-x}}{x}(1+\frac{1}{x})\nonumber\\ 
&&Y_2(x)=\frac{e^{-x}}{x}(1+\frac{3}{x}+\frac{3}{x^2})  \nonumber
\eea
For the 8-th component we have for the use with
non-strange baryons
\bea
A^0_{8, \pi}&\equiv& 0\\[.3cm]
{\bf A}_{8, \pi}&\equiv& {\bf 0} .
\eea
%%%%%%%%%%%%%%%%%%%%%%%%%%%%%%%%%%%%%%%%%%%%%%%%%%%%%%%%%%%%%%%%%%%%%%%
%WHY DO WE OBTAIN ZERO?
%:::::::::::::::::::::::::::::::::::::::::::::::::::::::::::::::::::::::::
% Can you explain this? A comment would improve the paper.
%:::::::::::::::::::::::::::::::::::::::::::::::::::::::::::::::::::::::::
%WHAT ABOUT THE AXIAL CURRENTS ASSOCIATED WITH ETA-EXCHANGE? 
%     f_{8kl} is different from zero only if k, l are 4,5,6,7. That means
%     you exchange kaons, but that is not possible between u and d quarks
%%%%%%%%%%%%%%%%%%%%%%%%%%%%%%%%%%%%%%%%%%%%%%%%%%%%%%%%%%%%%%%%%%%%%%%%

%As in Ref.\cite{david} $g_{Aq}$ is supplemented by a form factor as given
%by axial-vector meson dominance in Eq.(\ref{avmd})

\section{Spin content of the nucleon}
\label{section:results}
Because the spin operator of Dirac particles is related to the axial current
operator we can obtain information concerning the spin content of the nucleon
from the axial form factors calculated in the present chiral quark model. 
As explained in sect.~\ref{section:dis}, the spin fractions can be expressed in terms
of matrix elements of the axial current operators. 
%%%%%%%%%%%%%%%%%%%%%%%%%%%%%%%%%%%%%%%%%%%%%%%%%%%%%%%%%%%%%%%%%%%%%%%%%%%%%%%%%% 
%In the $\chi QM$,  the nucleon is made out of quarks and its  spin 
%is accounted for completely by the spin of the quarks and by their orbital
%angular momentum. Since the late $80's$ we know from the results of highly
%inelastic scattering on polarized protons that only a small fraction
%of the spin of the proton is carried by QCD quarks and that there is a
%significant contribution from the strange quarks in the Dirac sea \cite{emc}.
%More recent experiments conclude that approximately $1/3$ of the spin of the
%nucleon is carried by QCD quarks \cite{e143}. Calling $\Delta q$ the fraction
%of the proton spin carried by quarks of type $q$ we have from Ref.~\cite{e143}
%\bea
%\label{deltaq}
%\Delta u &=& \hspace{.35cm} 0.84 \pm 0.02 \nonumber\\
%\Delta d &=&-0.42 \pm 0.02 \nonumber\\
%\Delta s &=& -0.09 \pm 0.02. 
%\eea
%%%%%%%%%%%%%%%%%%%%%%%%%%%%%%%%%%%%%%%%%%%%%%%%%%%%%%%%%%%%%%%%%%%%%%%%%%%%%%%%%%%%
The latter 
are based on the larger SU(3)$_V \times$  SU(3)$_A$ symmetry underlying the
Lagrangian of the non-linear $\sigma$ model, which contains SU(3) flavor symmetry 
as a subgroup. There are now three axial vector couplings of the nucleon:
the flavor octet isovector axial coupling  $g_A(0)$, the flavor octet isosinglet 
(hypercharge) axial vector coupling $g_A^8(0)$, and the flavor singlet
axial coupling $g_A^0(0)$. 

\subsection{The flavor octet isovector axial coupling ${\bf g_A(0)}$}

The nucleon isovector axial coupling $g_A(0)$ can be obtained from the 
spin quantities defined above as
\bea
g_A(0)=\Delta u -\Delta d =1.26\pm 0.04.
\eea
This value agrees within errors with the one measured in neutron 
$\beta$-decay given by \hbox{$g_A^{n \to p}(0)=1.2670(35)$}. 
In Table~\ref{table:isovector} 
we give the results that we obtain in our model. The impulse contribution 
clearly dominates. Exchange currents account in this case for less than $9\%$ of 
the total value which is in reasonable agreement with the experimental number. 
As explained in Ref.~\cite{david} the fact that the axial coupling of the
constituents quarks $g_{Aq}$ is smaller than unity is mainly responsible 
for the improvement of the calculated $g_A(0)$ compared to the naive quark 
model result of $g_A(0)=5/3$. 

\subsection{The flavor octet isospin singlet axial coupling  ${\bf g_A^8(0)}$}
The flavor octet isospin singlet axial coupling $g_A^8(0)$
gives the fraction of the proton spin that is carried by valence QCD quarks if
the Dirac sea were SU(3) flavor symmetric.
Experimentally it is determined as
\bea
g_A^8(0)=\Delta u+\Delta d- 2 \Delta s =0.60\pm 0.08.
\eea

We evaluate $g_A^8(0)$ using the 8-th component of our axial
operators given in sect.~\ref{section:currents} and the nucleon wave function given in
sect.~\ref{section:hamiltonian}. 
From there we see that only impulse and gluon exchange currents 
contribute to $g_A^8(q^2)$ at four-momentum transfer $q^2=0$, 
which for nucleons also implies ${\bf q}^2=0$. 

In the first line of table 4 
we show our results with $g_{Aq}=1$. The impulse contribution is then very close to 1. 
This has to be so since ${\bf A}_{8, imp}({\bf q}^2=0)$ is $g_{Aq}$ times the total
spin operator $\sum_k \b{\sigma}_k$ and the quarks are mainly in relative 
S-wave states. 

Two-body exchange currents play a very important role for this observable.
The axial gluon exchange current, which effectively describes the $q {\bar q}$ 
pairs in the nucleon reduces the value obtained in impulse approximation 
by some $40\%$. In addition, there is a reduction of 
$g_A^8(0)$ because $g_{Aq} < 1$. The total  result is shown in the 
second line in Table~\ref{table:flavor}. Our total value is some $30\%$ smaller than the present 
experimental data. However, because SU(3) flavor symmetry is broken 
the condition $g^8_{Aq}=g_{Aq}$ need not be exact. Explicit
calculation of axial constituent quark couplings in the Nambu-Jona Lasinio model
shows that $g^8_{Aq} \ne g_{Aq}$~\cite{Yab93}. In summary, the experimentally observed 
spin fraction carried by the valence QCD quarks can be obtained 
if axial gluon exchange currents are included in the nonrelativistic quark model.

\subsection{The flavor singlet axial coupling ${\bf g_A^0(0)}$}
%:::::::::::::::::::::::::::::::::::::::::::::::::::::::::::::::::::
% REMOVED ``...of the nucleon''
%:::::::::::::::::::::::::::::::::::::::::::::::::::::::::::::::::::::

The other independent quantity that one can construct from the 
$\Delta q\, 's$ is $g_A^0(0)$ which is experimentally determined as
\bea
g_A^0(0)=\Delta u+\Delta d+\Delta s=0.33\pm 0.06.
\eea
The singlet axial coupling $g_A^0(0)$ is the nucleon spin fraction carried by all quarks, 
i.e., valence plus sea quarks.
%:::::::::::::::::::::::::::::::::::::::::::::::::::::::::::::::::::
% small change in wording 
%:::::::::::::::::::::::::::::::::::::::::::::::::::::::::::::::::::::

In order to evaluate $g_A^0(0)$ in our model we have to extend  chiral
symmetry to the group $U(3)_V\times U(3)_A$. 
The QCD Lagrangian is trivially invariant under $U(1)_V$ which is
the symmetry associated with baryon number conservation. 
As for $U(1)_A$ even though the QCD Lagrangian is invariant in the chiral limit 
we know that this symmetry is broken by quantum effects. This is 
called the $U(1)_A$ anomaly \cite{aa}. If the anomaly were not present
we would have another Goldstone boson $\Phi_0$ in the chiral symmetry limit $m_q \to 0$. 
However, the presence of the anomaly makes $m_{\Phi_0}$ nonzero even in 
the chiral limit of zero current quark masses. The natural candidate for this extra 
boson is the physical $\eta'$ meson whose mass is much larger than the masses 
of the pseudoscalar octet mesons, which are the quasi-Goldstone bosons of chiral 
$SU(3)$. In the large $N_C$ limit of QCD the $U(1)_A$ anomaly is absent. In that limit the 
massless QCD Lagrangian has the larger symmetry $U(3)_V\times U(3)_A$
and there are nine Goldstone bosons associated with the spontaneous 
chiral symmetry breaking to the subgroup $U(3)_V$.

In that large $N_C$ limit the Goldstone boson dynamics can be described 
 in terms of the matrix field 
\bea
 e^{i\lambda_0\Phi_0/f_{\pi}}\ \Sigma
\eea
where $\lambda_0=\sqrt{2/3}\, I$ and $\Sigma$ is given in Eq.(\ref{sigma}). 
The new boson field, $\Phi_0$, is invariant under
$U(3)_V\times SU(3)_A$ whereas under $U(1)_A$ it transforms as
\bea
\Phi_0 \stackrel{\stackrel{U(1)_A}{ }}{\longrightarrow} \Phi_0+a_0 f_{\pi}.
\eea
Here, $a_0$ is the parameter associated with the U(1)$_A$ transformation.
As for the other boson and quark fields the matrix field $\xi$ defined in
Eq.(\ref{xi}) is invariant
under U(1)$_V\times$ U(1)$_A$ and the quark field $\Psi$ transforms
under U(1)$_V$ as
\bea
\Psi \stackrel{\stackrel{U(1)_V}{ }}{\longrightarrow} e^{-i\lambda_0 a_0/2}\, 
\Psi
\eea 
being invariant under U(1)$_A$ because the $U$ matrix field does not
depend on $a_0$ or $\Phi_0$.

With these transformation properties, the most general Lagrangian
invariant under $U(3)_V\times U(3)_A$ has as the first few terms
\bea
\label{lag0}
&&{\cal L}=\overline{\Psi}i\gamma^{\mu}(\partial_{\mu}+{\cal V}_{\mu})
\Psi -M \overline{\Psi}\Psi+ g_{Aq}\overline{\Psi}\gamma^{\mu}\gamma_5{\cal A}_{\mu}
\Psi
+\frac{f^2_{\pi}}{4}\, tr\left( \partial^{\mu}\Sigma^{\dagger}
\partial_{\mu}\Sigma\right)\nonumber\\
&& \hspace{1cm}+\frac{g^0_{Aq}}{2f_{\pi}}\,\partial_{\mu}\Phi_0\,
\overline{\Psi}
\gamma^{\mu}\gamma_5\lambda_0\Psi
+\frac{1}{2}\,\partial_{\mu}\Phi_0\,\partial^{\mu}\Phi_0
+i\,g^0\,\frac{f_{\pi}}{2}\,\sqrt{\frac{2}{3}}\,\partial_{\mu}\Phi_0
\,tr\left(\Sigma\partial^{\mu}\Sigma^{\dagger}\right)
\eea
where $g^0_{Aq}$ and $g^0$ are free parameters not fixed by chiral symmetry
requirements alone.

Including also the current quark mass breaking term, where 
now the expression in 
 Eq.(\ref{cqmt}) is changed to include a factor $e^{-i\gamma_5\lambda_0\,\Phi_0
 /2f_{\pi}}$ to the left and to the right of $M_0$, 
and making an expansion 
 in powers of $1/f_{\pi}$ we get to dominant order 
\bea
\label{lag02}
{\cal L}=&&\hspace{-.5cm}\overline{\Psi}(i\gamma^{\mu}\partial_{\mu}-M-M_0)
\Psi\nonumber\\
%\hspace{-.2cm}  + \hspace{-.2cm}
&+& \frac{1}{2}\partial^{\mu}\Phi_j
\partial_{\mu}\Phi_j-\frac{1}{2}\,m^2_{\Phi_j}\,\Phi^2_j
+ \frac{1}{2}\partial^{\mu}\Phi_0
\partial_{\mu}\Phi_0-\frac{1}{2}\,m^2_{\Phi_0}\,\Phi^2_0
-m^2_{\Phi_{8,0}}\,\Phi_8\Phi_0
\nonumber\\
&+&\hspace{-.2cm} \frac{g_{Aq}}{2f_{\pi}}\,\overline{\Psi}\gamma^{\mu}\gamma_5\lambda_j
\Psi\partial_{\mu}\Phi_j
+ \frac{g^0_{Aq}}{2f_{\pi}}\,\overline{\Psi}\gamma^{\mu}\gamma_5\lambda_0
\Psi\partial_{\mu}\Phi_0
+\frac{i}{2f_{\pi}}\, \overline{\Psi}\gamma_5\{\lambda_{\alpha},M_0\}
\Psi\, \Phi_{\alpha}\nonumber\\
&-&\frac{1}{4f^2_{\pi}}\,f_{jkl}\overline{\Psi}
\gamma^{\mu}\lambda_j\Psi\,\Phi_k \partial_{\mu}\Phi_l
+\frac{1}{8f^2_{\pi}}\,\overline{\Psi}\{\lambda_{\alpha},\{\lambda_{\beta},M_0
\}\}\,\Psi\, \Phi_{\alpha}\Phi_{\beta}
\eea
where the $\alpha$ and $\beta$ indices run from 0 to 8. 

The mass term for the $\Phi_0$ meson has to be supplemented by a quantity that
is not vanishing in the chiral limit of zero quark masses, see for instance
Ref.~\cite{pich}. 
Due to the $\Phi_8$ and  $\Phi_0$ mass mixing term  those bosons 
are no longer the physical states. The physical states are the 
$\eta$ and $\eta'$ mesons obtained by diagonalizing the mass matrix and that are given in terms of   
 $\Phi_8$ and $\Phi_0$ as
\bea
\eta &=& \cos \theta_P\, \Phi_8-\sin \theta_P\, \Phi_1\nonumber\\
\eta' &=& \sin \theta_P\, \Phi_8+\cos \theta_P\, \Phi_1
\eea
where $\theta_P$ is the pseudoscalar mixing angle. There are different 
determinations of  $\theta_P$ in the literature. In the theoretical analysis of
Ref.\cite{feldmann} a value of $\theta_P=-12.3^o$ is obtained while a 
phenomenological analysis of different experimental data done by the same
 authors gives \hbox{$\theta_P=-15.4\pm 1^o$}. In Ref.\cite{bramon} the study
of vector-pseudoscalar-photon decays gives \hbox{$\theta_P=-17\pm 2.9^o$}.
A recent evaluation in Lattice QCD done in Ref.\cite{ukqcd}
gives the result \hbox{$\theta_P=-10\pm 2^o$}. Finally the value 
 favored by the Particle Data Group \cite{pdg} is
\hbox{$\theta_P \approx -20^o$} which is obtained
from the analysis of pseudo-scalars two-photon decays. Different determinations
  range from 
$-10^o$ to $-20^o$.

Turning now to currents it is easy to see that for the vector currents
with components \hbox{ $j=1,2,3,8$} 
there is no difference from the results stated in 
Eqs.(\ref{vectorc},\ref{divervec}).
The new $U(1)_V$ symmetry has an associated vector current that is given by
\bea
V^{\mu}_0=\overline{\Psi}\gamma^{\mu}\frac{\lambda_0}{2}
\Psi .
\eea
Apart from a constant factor, this is the baryon number current and is
exactly conserved
\bea
\partial_{\mu}\,V^{\mu}_0=0 .
\eea

For the axial currents, and compared to the results in sect.~\ref{section:sigma},  we have 
that the divergence of the  $j=8$ component is now given by
\bea
\label{pcac1}
\partial_{\mu}\, A_8^{non-pole,\, \mu}\equiv -\sqrt{2}\,f_{\pi}\left\{\cos \theta_P
\left(\partial_{\mu}\,\partial^{\mu}+m^2_{\eta}\right) \eta
+\sin \theta_P
\left(\partial_{\mu}\,\partial^{\mu}+m^2_{\eta'}\right) \eta'\right\}
\eea
while for the $j=1,2,3$ components there is no change.
As for the new axial current corresponding to the  $U(1)_A$ symmetry
we have for the non-pole part 
\bea
\label{a0}
A^{non-pole,\, \mu}_0=g^0_{Aq}\,\overline{\Psi}\gamma^{\mu}
\gamma_5\frac{\lambda_0}{\sqrt{2}}
\Psi
\eea
whose divergence is given by
\bea
\label{pcac2}
\partial_{\mu}\, A_0^{non-pole,\, \mu}\equiv -\sqrt{2}\,f_{\pi}\left\{ -\sin \theta_P
\left(\partial_{\mu}\,\partial^{\mu}+m^2_{\eta}\right) \eta
+\cos \theta_P \left(\partial_{\mu}\,\partial^{\mu}+m^2_{\eta'}\right) \eta'
\right\} .
\eea

The two PCAC relations in Eqs.(\ref{pcac1},\ref{pcac2}) translate into the 
following relations in momentum space
\bea
\label{axial80}
q_{\mu}\ A^{non-pole,\,\mu}_8&=& i\ \sqrt2\ f_{\pi}\bigg\{ \hspace{.4cm}\cos \theta_P\ M^{\eta}
+ \sin \theta_P\ M^{\eta'} \bigg\} \nonumber\\
q_{\mu}\ A^{non-pole,\,\mu}_0&=& i\ \sqrt2\ f_{\pi}\bigg\{ -\sin \theta_P\ M^{\eta}
+\cos \theta_P\ M^{\eta'} \bigg\}
\eea
where  $A^{non-pole,\,\mu}$ stands here for the non-pole part of the axial current
amplitude and $M^{\eta,\eta'}$ is the
absorption amplitude for the $\eta,\eta'$ mesons. Relations in Eq.(\ref{axial80}) are exactly satisfied for each independent contribution namely: 
impulse of Figs. 2-3a, one-gluon exchange of Figs. 2-3b and confinement 
exchange of \hbox{Figs. 2-3c.} 
%The coupling contants of the $\eta$ and $\eta'$
%to the $u$ and $d$ quarks are given by
%\bea
%\label{ee'cc}
%\frac{g_{\eta q}}{2m_q}&=&\frac{1}{2 f_{\pi}\sqrt3}\bigg(\cos \theta_P\, g_{Aq}
%-\sin \theta_P\,\sqrt2\,
% g^0_{Aq}\ \bigg)\nonumber \\
%\frac{g_{\eta' q}}{2m_q}&=&\frac{1}{2 f_{\pi}\sqrt3}\bigg(\sin \theta_P\, g_{Aq}
%+\cos \theta_P\,\sqrt2\,
% g^0_{Aq}\ \bigg).
%\eea

To evaluate $g_A^0(0)$ we will need the non-relativistic operators associated
with $A^{non-pole,\, \mu}_0$ current. From Eq.(\ref{a0}) and our
discussion in sect.~\ref{section:sigma} it is clear
that there will be impulse, gluon and confinement exchange current contributions
similar to the ones introduced in sect.~\ref{section:currents} 
for the 8-th component of the axial
current. In fact, with the correct normalization and in the case of non-strange
baryons, the operators are exactly the same as those given in sect.~\ref{section:currents} 
for the 
8-th component of the axial current with the sole replacement 
of $g_{Aq}$ by $g^0_{Aq}$.

Our results for $g_A^0(0)/g^0_{Aq}$  are  the same as in Table~\ref{table:flavor}
for $g_A^8(0)/g_{Aq}$. In order to reproduce the experimental value of
$g_A^0(0)=0.33$ we need
\bea
 g^0_{Aq}=0.553
\eea
which is $30\%$ smaller than $g_{Aq}$. The ratio
\bea
\zeta=g^0_{Aq}/g_{Aq}=0.714
%\approx 1/\sqrt2
\eea 
obtained here is at variance with other empirical determinations like 
$\vert \zeta \vert =1.16$ 
%%%%%%%%%%%%%%%%%%%%%%%%%%%%%%%%%%%%%%%%%%%%%%%%%
% HERE I NOTED 1.26  ?
%%%%%%%%%%%%%%%%%%%%%%%%%%%%%%%%%%%%%%%%%%%%%%%%
in 
Ref.\cite{glozman} or $\zeta=-1.2$ in Ref.\cite{cheng}. It is however, in
agreement with the result $g^0_{Aq}=0.54$ obtained in Ref.~\cite{Yab93}. 

A criticism is in order here. In our Hamiltonian we have used an $\eta_8$, 
or $\Phi_8$, 
meson exchange potential
with a $\eta_8$ coupling to $u$ and $d$ quarks  given by $g_{Aq}/(2 f_{\pi}\sqrt3)$.
With $\eta_0$ and $\eta_8$ mixing one would have to change the $\eta_8$ 
exchange potential into corresponding  $\eta$ and $\eta'$ exchange potentials 
with couplings to $u$ and $d$ quarks  given by
\bea
\label{ee'cc}
\frac{g_{\eta q}}{2m_q}&=&\frac{1}{2 f_{\pi}\sqrt3}\bigg(\cos \theta_P\, g_{Aq}
-\sin \theta_P\,\sqrt2\,
 g^0_{Aq}\ \bigg)\nonumber \\
\frac{g_{\eta' q}}{2m_q}&=&\frac{1}{2 f_{\pi}\sqrt3}\bigg(\sin \theta_P\, g_{Aq}
+\cos \theta_P\,\sqrt2\,
 g^0_{Aq}\ \bigg).
\eea
%::::::::::::::::::::::::::::::::::::::::::::::::::::::::::::::::::::::::::::::::
% Eq.(\ref{ee'cc}) no longer exists....
%:::::::::::::::::::::::::::::::::::::::::::::::::::::::::::::::::::::::::::::::::
While this change will affect
the determination of parameters and the nucleon wave function,
the fact that the $\eta_8$ and the $\eta$ and $\eta'$ meson masses are much larger 
than the pion mass makes the corresponding potentials  not
so relevant in this respect. Their 
effect can be compensated by very small changes in the free parameters of the model.

\subsection{Effective couplings of the $\eta$ and $\eta'$ mesons to nucleons}

Using PCAC at the baryon level one can get relations between $g^8_A(0)$ and
$g^0_A(0)$ on one side and the $\eta$ and $\eta'$ coupling to nucleons on the
other.
Those relations are
\bea
g^8_A(0)=\sqrt3\,f_{\pi}\left\{\hspace{0.4cm} \cos \theta_P \frac{g_{\eta NN}
(0)}{M_N}
+\sin \theta_P \frac{g_{\eta' NN}(0)}{M_N}\right\}
\eea
and similarly
\bea
g^0_A(0)=\frac{\sqrt3}{\sqrt2}\,f_{\pi}
\left\{ -\sin \theta_P \frac{g_{\eta NN}(0)}{M_N}
+\cos \theta_P \frac{g_{\eta' NN}(0)}{M_N}\right\}
\eea
from where the $\eta N N$ and $\eta' N N$ 
 coupling constants can be obtained as 
\bea
g_{\eta NN} &=& \frac{M_N}{\sqrt3\,f_{\pi}}\left\{\cos \theta_P\, g^8_A(0)-
\sin \theta_P\, \sqrt2\,g^0_A(0)\right\}\nonumber\\
g_{\eta' NN} &=& \frac{M_N}{\sqrt3\,f_{\pi}}\left\{\sin \theta_P\, g^8_A(0)+
\cos \theta_P\,\sqrt2\, g^0_A(0)\right\}.
\eea
Using our theoretical results for  $g_A^8(0)$ and   $g_A^0(0)$ (the second one
is in fact fixed to the experimental value), and with the use of
$\theta_P$ in the range $[-20^o,-10^o]$ we get
\bea
g_{\eta NN} &=& 3.18-3.52\nonumber\\
g_{\eta' NN} &=&2.22- 1.63
\eea
or
\bea
\alpha_{\eta NN}=\frac{g^2_{\eta NN}}{4\pi}=0.80-0.99\nonumber\\
\alpha_{\eta' NN}=\frac{g^2_{\eta' NN}}{4\pi}=0.39-0.21.
\eea

There are various determinations of $\alpha_{\eta NN}$
and $\alpha_{\eta' NN}$ in the literature. Typical values for
$\alpha_{\eta NN}$ suggested by the Bonn model \cite{bonnbm} 
are in the range $3-5$, much larger that our evaluation.
From a practical viewpoint one need not worry too much about 
this discrepancy as the $\eta$ meson is not very relevant for the 
fit of the nucleon-nucleon phase shifts. Also its
contribution to the nuclear binding is very small for normal nuclear densities.
Besides, in $NN$ potentials the $\eta NN$ coupling is usually
treated as an effective parameter in which contributions due to two-meson exchanges
are absorbed. The analysis of the $\pi^- p\to \eta n$ reaction in Ref.\cite{peng} 
results in values for $\alpha_{\eta NN}$ in the range $0.6-1.7$ in good agreement with
our determination. Smaller values are found in Ref.\cite{kirchbach} where the
analysis of the $a_0\to \eta \pi$ decay gives
$\alpha_{\eta NN}=0.315$. Also the use of QCD sum rules on the light 
cone~\cite{zhu} 
gives $\alpha_{\eta NN}=0.35\pm 0.1$.

As for $\alpha_{\eta' NN}$, a recent theoretical 
analysis of $p p$ scattering near the production threshold~\cite{moskal} 
gives $\alpha_{\eta' NN} < 0.5$. Another analysis~\cite{zhao}, 
this time of the near threshold $\eta'$ photo-production, 
obtains
$\alpha_{\eta' NN}\approx 0.22$. Both results are in good agreement with
our evaluation.

\section{Summary}
\label{section:summary}

We have investigated the axial form factors related to the spin structure 
of the nucleon in the framework of a chiral quark model 
where chiral symmetry is introduced via a non-linear $\sigma$ model.
We have included one-body and two-body axial currents as required by
the PCAC condition. For $g_A(0)$ we obtain a value of 1.34 in good agreement 
with experiment. We find the $g_A(0)$ is dominated by the one-body 
axial current. For $g_A^8(0)$, related to the spin fraction carried
by the valence quarks (if the sea is flavor symmetric), 
we obtain 0.47. This surprising result within the context of 
a nonrelativistic quark potential model, where one would expect a value of 1,
is due to two factors.  First, the inclusion of the relativistic correction
provided by the axial one-gluon exchange current, and second the fact that the
constituent quark axial coupling constant $g_{Aq}$ satisfies the Goldberger-Treiman
relation of Eq.(\ref{gaq}) and is therefore smaller than unity.
 Due to the internal structure of a constituent quark, also its 
coupling to the singlet axial current $g^0_{Aq}$ gets renormalized. 
Its value is considerably smaller 
than unity. 

Using our predictions for $g_A^8(0)$ and $g_A^0(0)$ as well as the PCAC 
constraint, we calculate the effective 
coupling constants of the $\eta$ and $\eta'$ mesons to nucleons. The values 
obtained for these coupling constants are in agreement with different
theoretical analyses done by other groups.

\noindent
{\bf ACKNOWLEDGMENTS:}\\
Work supported in part 
by Spanish DGICYT under contract no. BFM2000-1326 and Junta de Castilla
y Leon under contract no. SA109/01

\newpage

\begin{table}
\caption{Quark model parameters: $b$ is the harmonic oscillator constant,
$\alpha_s$ is the quark-gluon coupling strength, $a$
is the confinement strength, $\mu$ the color screening
length, and $C$ a constant term in the color screening
confinement potential.}
%\vspace{2.0cm}
\label{table:parameters}
\begin{center}
\nobreak
\begin{tabular}[t]{ c    c  c  c c} 
%\hline
   $b [fm]$ 
 &  $\alpha_s$ &
 $a [MeV]$ &  $\mu [fm^{-1}]$ &  $C [MeV]$
\\ 
\hline   
%\hline
 & & & & \\
 0.591 & 0.987  & 244.041  & 0.770 & -279.881  
                                           \\                   
%\hline  
\end{tabular}
\end{center}
\end{table}

\noindent
\begin{table}
\caption{ Admixture coefficients in the wave function of 
Eq.(\protect{\ref{wf}}),
evaluated with the Hamiltonian of 
Eq.(\protect{\ref{hamiltonian}}) 
 in an $N=2$ harmonic oscillator space.}
%\vspace{2.0cm}
\label{table:admixtures}
\begin{center}
\nobreak
\begin{tabular}[t]{ c  c  c c  c } 
%\hline
  $a_{S_S}$ &  $a_{S'_S}$ 
 &  $a_{S_M}$ &
 $a_{D_M}$ & $a_{P_A}$
\\ 
\hline  
 & & & & \\ 
%\hline
%
0.9585      & -0.1475  & -0.2344  & -0.0672 & 0.0011   
                                             \\                   
%\hline  
\end{tabular}
\end{center}
\end{table}

%:::::::::::::::::::::::::::::::::::::::::::::::::::::::::::::::::::::::::::::::::::::
% I think it is important to include the experimental numbers in the table .
%:::::::::::::::::::::::::::::::::::::::::::::::::::::::::::::::::::::::::::::::::::::

\noindent
\begin{table}
\caption{ Results for $g_A(0)$.
Imp., impulse contribution ; g, gluon exchange contribution ; 
$\pi$, pion exchange contribution ; Conf., 
confinement contribution; Total, total result.}
%\vspace{2.0cm}
\label{table:isovector}
\begin{center}
\nobreak
\begin{tabular}[t]{ c  c c c  c c c c c} 
%\hline
  & Imp. & g &  $\pi$   & Conf.& Total & Exp.
\\ 
\hline 
& & & & &  &\\ 
& & & & &  & \\
$g_A(0)$  & 1.228 & -0.206 & 0.321  & 0 & 1.343 & 1.2670(35)\\  
 & & & & & &  \\
%\hline 
\end{tabular}
\end{center}
\end{table}

%:::::::::::::::::::::::::::::::::::::::::::::::::::::::::::::::::::::::::::::::::::::
% I think it is important to include our result for g_A^0 in the table.
% and also the experimental numbers .
%:::::::::::::::::::::::::::::::::::::::::::::::::::::::::::::::::::::::::::::::::::::
%

\noindent
\begin{table}
\caption{ Results for $g_A^8(0)$ and $g_A^0(0)$.
Imp., impulse contribution; g, gluon exchange contribution; Total,
 total result. The constituent quark axial couplings are given by
$g_{Aq}=g^8_{Aq}=0.774$ and $g^0_{Aq}=0.553$.  }
\label{table:flavor}
\begin{center}
\nobreak
\begin{tabular}[t]{ c  c  c c c } 
%\hline
  & Imp.  & g & Total & Exp.
\\ 
\hline 
& & & \\
$g_A^8(0)/g_{Aq}$  & 0.9909 & -0.3861 & 0.6048\\   
& & & & \\
$g_A^8(0)$  & 0.7670 & -0.2988 & 0.4682 & 0.60(8)\\   
& & & &  \\
$g_A^0(0)/g^0_{Aq}$  & 0.9909 & -0.3861 & 0.6048\\ 
& & & & \\
$g_A^0(0)$  & 0.5480 & -0.2135 & 0.3345 & 0.33(6)\\   
%\hline 
\end{tabular}
\end{center}
\end{table}

\newpage

\begin{figure}
\begin{center}
\rotatebox{0}{\resizebox{14cm}{6cm}{\includegraphics{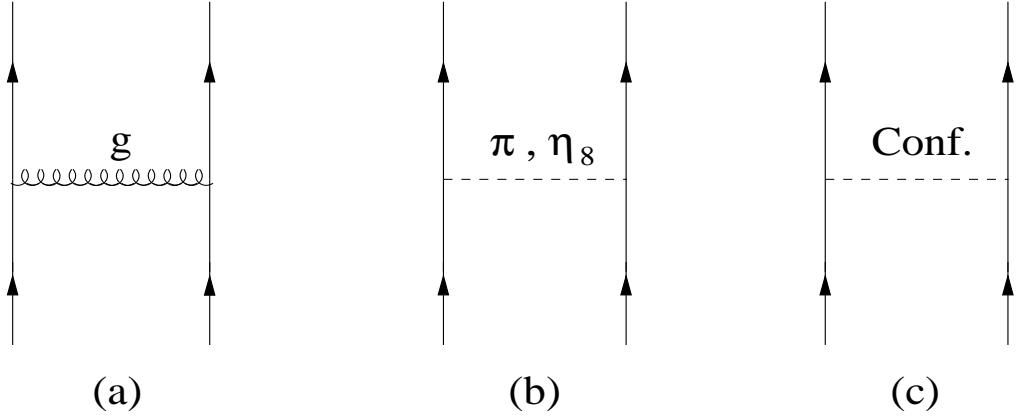}}}
\parbox[b]{13cm}{\caption{Feynman diagrams for two-body exchange potentials.
a) One-gluon exchange potential. b) One-pion and one-$\eta_8$ exchange 
potential. c) Confinement potential.}}
\end{center}
\end{figure}

\begin{figure}
\begin{center}
\rotatebox{0}{\resizebox{17cm}{12cm}{\includegraphics{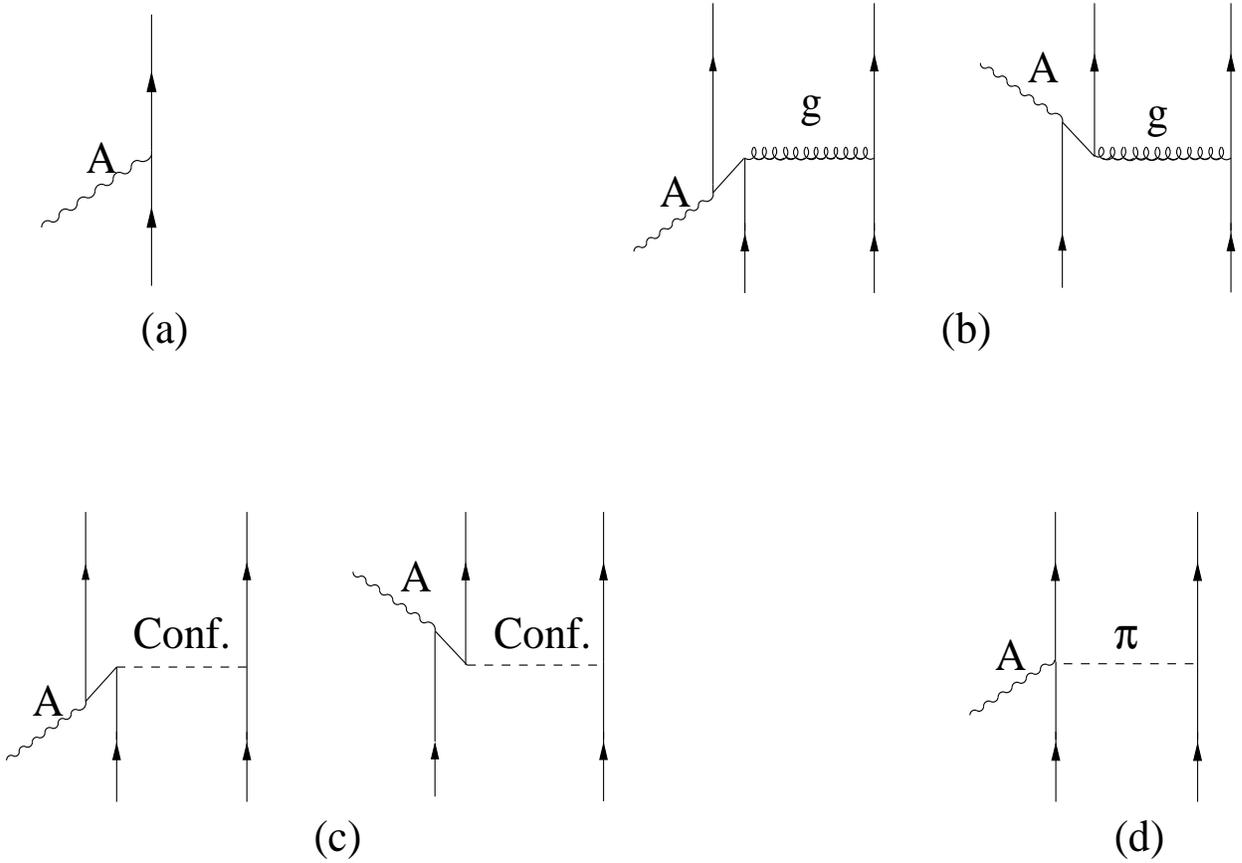}}}
\parbox[b]{13cm}{\caption{Feynman diagrams for the axial operators.
a) Impulse. b) One-gluon exchange. c) Confinement interaction.
d) Pion exchange.}}
\end{center}
\end{figure}

\begin{figure}
\begin{center}
\rotatebox{0}{\resizebox{17cm}{12cm}{\includegraphics{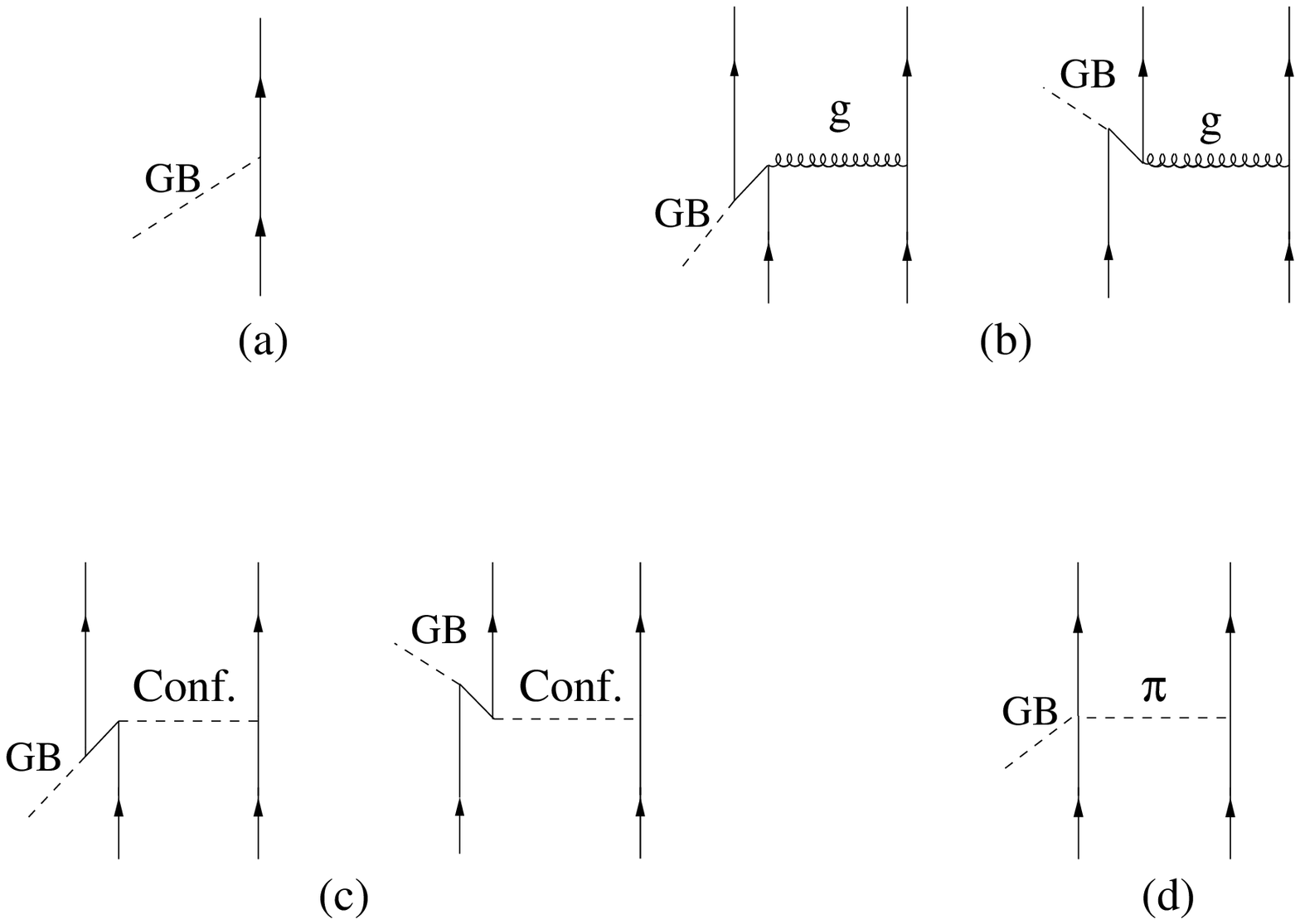}}}
\parbox[b]{13cm}{\caption{Feynman diagrams for the Goldstone boson (GB) 
absorption operators.
a) Impulse. b) One-gluon exchange. c) Confinement exchange.
d) Pion exchange.}}
\end{center}
\end{figure}


\begin{thebibliography}{99}
\bibitem{david} D. Barquilla-Cano, A.J. Buchmann 
and E. Hern\'andez, Nucl. Phys. {\bf A714}, 611 (2003).
%\bibitem{Ada91} J. Adam, Ch. Hajduk, H. Henning, P. U. Sauer, E. Truhlik,
%Nucl. Phys. {\bf A531}, 623 (1991). 
\bibitem{emc} J. Ashman et al. (European Muon Collaboration),
Phys. Letts. B {\bf 206}, 364 (1988); Nucl. Phys. {\bf B328}, 1 (1989).
\bibitem{abe} K. Abe et al. (E143 Collaboration), Phys. Rev. D {\bf 58},
112003 (1998)
\bibitem{carlitz} R. D. Carlitz, Int. J. Mod. Phys. E{\bf 1}, 505 (1992).
\bibitem{lipkin} H. J. Lipkin, Phys. Lett. B {\bf 214}, 429 (1988).
\bibitem{flores} R. Flores-Mendieta, E. Jenkins and A.V. Manohar, 
Phys. Rev. D {\bf 58}, 094028 (1996).
\bibitem{jackson} H. E. Jackson, Int. J. Mod. Phys. A{\bf 17} (2002) 3551.
See also the talk by H. E. Jackson in the present proceedings.
\bibitem{leader} E. Leader and D. B. Stamenov, Phys. Rev. D {\bf 67}, 037503
(2003)
\bibitem{manohar} A. Manohar and H. Georgi, Nucl. Phys. {\bf B234}, 189 (1984). 
%\bibitem{vento} J. Navarro and V. Vento, Phys. Lett. B {\bf 140},6 (1984); 
%V. Vento and J. Navarro, Phys. Lett. B {\bf 141}, 28 (1984). 
\bibitem{w1} S. Weinberg, Phys. Rev. Lett. {\bf 65}, 1181 (1990). 
\bibitem{w2}S. Weinberg, Phys. Rev. Lett. {\bf 67}, 3473 (1991). 
\bibitem{andreas} W. Broniowski, M. Lutz and A. Steiner, Phys. Rev. Lett.
{\bf 71}, 1787 (1993). 
\bibitem{vogl} U. Vogl, M. Lutz, S.Klimt and  W. Weise, Nucl. Phys. {\bf A516},
469 (1990). 
\bibitem{weinbergbook} S. Weinberg, ``The Quantum Theory of Fields'',
Cambridge University Press 1966, Vol II, pags.228-230.
\bibitem{paco} F. Fern\'andez and E. Oset, Nucl. Phys.  {\bf A455}, 720 (1986).
\bibitem{weise} W. Weise, Nucl. Phys. {\bf A434}, 685c (1985). 
\bibitem{rujula} A. de R\'ujula, H. Georgi and S.L. Glashow, Phys. Rev. D {\bf 12}, 147 
(1975).
\bibitem{ik} N. Isgur and G. Karl, Phys. Rev. D {\bf 18}, 4187 (1978).
\bibitem{ik2} N. Isgur, G. Karl and R. Koniuk, Phys. Rev. Lett. {\bf 41},
1269 (1978); N. Isgur and G. Karl, Phys. Rev. D {\bf 19}, 2653 (1979).
\bibitem{meyer}U. Meyer, E. Hern\'andez and A. J. Buchmann, Phys. Rev. C{\bf 64}, 
035203 (2001).
\bibitem{Yab93} H. Yabu, M. Takizawa, W. Weise, Z. Phys. A {\bf 345}, 193 (1993).  
\bibitem{aa} S. L. Adler, Phys. Rev. {\bf 177}, 2426 (1969);
J. S. Bell and R. Jackiw, Nuo. Cim. {\bf A60}, 47 (1969);
S. L. Adler and W. A. Bardeen, Phys. Rev. {\bf 182}, 1517 (1969).
%\bibitem{gari} M. Gari and U. Kaulfuss, Phys. Lett. B {\bf 138}, 29 (1984).
%\bibitem{pdg} D.E. Groom et al. (Particle Data Group), Eur. Phys. J. C
%{\bf 15}, 1 (2000). 
%\bibitem{liesenfeld} A. Liesenfeld et al., Phys. Lett. B {\bf 468}, 19 (1999). 
%\bibitem{go} M. Gell-Mann, Phys. Rev. {\bf 106}, 1296 (1957);
%S. Okubo, Prog. Theor. Phys. {\bf 27}, 949 (1962).
%\bibitem{zhang} Z.-Y. Zhang, Y.-W. Yu, P.-N. Shen,
%X.-Y. Shen and Y.-B. Dong, Nucl. Phys. {\bf A561}, 595 (1993).
%\bibitem{weinberg} S. Weinberg, ``The Quantum Theory of Fields'' (Cambridge
%University Press 1996), Vol II, pag. 228.
%\bibitem{Yab93} H. Yabu, M. Takizawa, W. Weise, Z. Phys. A {\bf 345}, 193 (1993).  
\bibitem{pich} A. Pich, Rept. Prog. Phys. {\bf 58}, 563 (1995)
\bibitem{feldmann} Th. Feldmann, P. Kroll and B. Stech, Phys. Rev. D {\bf 58},
114006 (1998).
\bibitem{bramon} A. Bramon, R. Escribano and M. D. Scadron, Eur. Phys. J.
C {\bf 7}, 271 (1999).
\bibitem{ukqcd} C. McNeile and C. Michael (UKQCD Collaboration),
Phys. Lett. B {\bf 491}, 123 (2000).
\bibitem{pdg} D.E. Groom et al. (Particle Data Group), Eur. Phys. J. C
{\bf 15}, 1 (2000). 
\bibitem{glozman} L. Ya. Glozman, W. Plessas, K. Varga and R.F Wagenbrunn,
Phys. Rev. D {\bf 58}, 094030 (1998).
\bibitem{cheng} T. P. Cheng and L.-F. Li, Phys. Rev. Lett. {\bf 74},
2872 (1995).
\bibitem{bonnbm} R. Brockmann and R. Machleidt, Phys. Rev. C {\bf 42},
1965 (1990).
\bibitem{peng} J. C. Peng, Proc. of the LAMPF Workshop on Photon and Neutral
Meson Physics at Intermediate Energies-LA-11177-C, Ed. H. W. Baer et al, 1987.
\bibitem{kirchbach} M. Kirchbach and L. Tiator, Nucl. Phys. {\bf A604},
385 (1996).
\bibitem{zhu} S.-L. Zhu, Phys. Rev. C {\bf 61}, 065205 (2000). 
\bibitem{moskal} P. Moskal et al., Phys. Rev. Lett. {\bf 80}, 3202 (1998).
\bibitem{zhao} Q. Zhao, Phys. Rev. C {\bf 63}, 035205 (2001).
\end{thebibliography}
\end{document}